\documentclass[aps,superscriptaddress,showpacs,nofootinbib,eqsecnum]{revtex4}
 
\usepackage{epsfig} 
\input{epsf}
 
\begin{document} 
 
\title{Critical and Tricritical Points for the Massless 2d Gross-Neveu Model
  Beyond large-$N$}
 
\author{Jean-Lo\"{\i}c Kneur} \email{kneur@lpta.univ-montp2.fr}
\affiliation{Laboratoire de Physique Math\'{e}matique et Th\'{e}orique - CNRS
  - UMR 5825 Universit\'{e} Montpellier II, France}

\author{Marcus Benghi Pinto} \email{marcus@fsc.ufsc.br}
\affiliation{Departamento de F\'{\i}sica, Universidade Federal de Santa
  Catarina, 88040-900 Florian\'{o}polis, Santa Catarina, Brazil}
 
\author{Rudnei O. Ramos} \email{rudnei@uerj.br} \affiliation{Departamento de
  F\'{\i}sica Te\'orica, Universidade do Estado do Rio de Janeiro, 20550-013
  Rio de Janeiro, RJ, Brazil}

\begin{abstract} 
  Using optimized perturbation theory, we evaluate the effective potential for
  the massless two dimensional Gross-Neveu model at finite temperature and
  density containing corrections beyond the leading large-$N$ contribution.
  {}For large-$N$, our results exactly reproduce the well known $1/N$ leading
  order results for the critical temperature, chemical potential and
  tricritical points.  {}For finite $N$, our critical values are smaller than
  the ones predicted by the large-$N$ approximation and seem to observe
  Landau's theorem for phase transitions in one space dimension. New
  analytical results are presented for the tricritical points that include
  $1/N$ corrections.  The easiness with which the calculations and
  renormalization are carried out allied to the seemingly convergent optimized
  results displayed, in this particular application, show the robustness of
  this method and allows us to obtain neat analytical expressions for the
  critical as well as tricritical values beyond the results currently known.
 
\end{abstract} 
 
 
\maketitle
 
\section{Introduction}

The study of symmetry breaking/restoration in quantum field theories is an
important issue of relevance in many areas of physics.  {}For example, today,
problems regarding phase transitions in Bose-Einstein condensates (BEC) or in
Quantum Chromodynamics (QCD) concentrate a lot of theoretical as well as
experimental efforts.
 
Of topical importance regarding studies of phase transitions in quantum field
theory is the reliability of perturbation theory and its eventual breakdown.
{}For instance perturbation theory at high temperatures breaks down due to the
appearance of large infrared divergences, happening for example in massless
field theories, like in QCD \cite{gross}, close to critical temperatures (in
field theories displaying a second order phase transition or a weakly first
order transition \cite{GR,beyond}), or just because that at high temperatures
there are parameter regimes where conventional perturbation schemes become
unreliable when powers of the coupling constants become surmounted by powers
of the temperature.  In these cases, a nontrivial problem arises since
non-perturbative methods must be used.  Various non-perturbative techniques
have been used to deal with these problems. Among them we will be particularly
interested in the $1/N$ approximation \cite{largeNreview} which, here, will be
considered mainly for comparison.  Though a powerful resummation method, the
$1/N$ approximation can quickly become cumbersome after the resummation of the
first leading contributions, like for example in $N=2$ (e.g, BEC and
polyacetylene) or $N=3$ (e.g., QCD) finite $N$ problems.  This is due to
technical difficulties such as the formal resummation of infinite subsets of
{}Feynman graphs and their subsequent renormalization.
 
An alternative non-perturbative analytical method that we will make use in
this work is the optimized perturbation theory, or linear $\delta$ expansion
(LDE) \cite{linear}.  In calculational terms (including renormalization) its
appeal regards the fact that one remains within the framework of perturbation
theory. Then, non-perturbative results are obtained by optimizing the
perturbatively evaluated quantities. This procedure amounts to eliminate,
variationally, mass parameters used to deform (interpolate) the original
action. Recently, the LDE has been successful in treating scalar field
theories at finite temperature and/or density.  Relativistic scalar $\lambda
\phi^4$ theories have been treated at finite temperature \cite{prd1} as well
as at finite temperature and density \cite{hugh}. At the same time, their
nonrelativistic counter part, which is relevant for BEC, has been studied in
connection with the problem of the dependence with interactions of the
critical temperature shift $\Delta T_c$, given by the difference between the
interacting and ideal gas critical temperatures (for a recent review see for
instance \cite{shiftTc} and references therein).  It suffices to say that the
latest LDE results \cite{knp} are in excellent agreement with the Monte Carlo
results, considered the most accurate prediction for $\Delta T_c$, and it
performs much better than the $1/N$ expansion used at leading and at next to
the leading order (for a review on the different results and methods used to
study the $\Delta T_c$ in the BEC problem, see \cite{jeans}).  Moreover, the
convergence properties for this critical theory have been proved by the
present and other authors \cite{new,knp,braaten}.
 
The present work will focus on the LDE applications to problems involving
phase transitions in asymptotically free models at finite temperatures and
densities. Here, we shall consider the 1+1 dimensional Gross-Neveu model (GN),
which is extensively used as a prototype model in studies related to phase
transitions in particle physics as well as condensed matter physics. One
recent application concerning particle physics is the next-to-leading order in
$1/N$ evaluation of the effective potential at finite temperature performed in
Ref.  \cite{blaizot}. In the condensed matter domain, a very interesting
recent application of the GN model was to the study of polymers
\cite{gnpolymers}, where a {\it massive} version of the model was used to show
the appearance of a kink-antikink crystal phase that was missed in a previous
work \cite{italianos}.

In Ref. \cite{gnpolymers} the authors work only within the leading order of
the 1/$N$ expansion, so their results still show a phase transition at finite
temperature $T$ and density $\mu$. On the other hand they consider the case of
inhomogeneous background fields, which should be physically more relevant in
the 1+1 dimensional GN model.  This is so because, in one space dimension, due
to the Landau-Mermin-Wagner-Coleman theorem \cite{landau,mermin,coleman1}, no
phase transition related to a discrete symmetry breaking (in this case a
discrete chiral symmetry in the massless GN model considered in this work) is
expected at any finite temperature. In the GN model this can be explained by
the role played by kink-like inhomogeneous configurations \cite{ma} that come
to dominate the action functional, instead of just homogeneous, constant field
configurations.  By accounting for kink-like configurations in the large-$N$
approximation, the authors of Ref.  \cite{gnpolymers} (see also Ref.
\cite{thiesreview} for a review) find evidence for a crystal phase that shows
up in the extreme $T \sim 0$ and large $\mu$ part of the phase diagram.  The
other extreme of the phase diagram, large $T$ and $\mu \sim 0$, remains
identical to the usual large-$N$ results for the critical temperature and
tricritical points, which are well known results \cite{wrongtc,muc,others}.
In the study performed here, we only consider homogeneous backgrounds, but go
beyond large-$N$, so the phase diagram changes as a whole. However, we cannot
see any crystal phase at small $\mu$, that should be a consequence of kink
like configurations dominating the action functional at that extreme of the
phase diagram.  Despite of that, we are still improving the calculations for
the GN model even though we are not considering inhomogeneous fields. And, as
explained above, since inhomogeneous backgrounds do not seem to change
appreciably the large $T$ and small $\mu$ region, up to the tricritical point
of the phase diagram, we are certainly improving the knowledge in that part of
the phase diagram. At the same time, we believe that our results are not
faithful in the small $T$ and large $\mu$ part of the phase diagram, which
gets affected at large by inhomogeneous backgrounds as shown by the results
obtained for the GN model in the large-$N$ approximation \cite{gnpolymers},
though no results beyond large-$N$ are currently available. Eventually, in the
future it would be opportune to contrast the results found by the authors of
Ref.  \cite{gnpolymers} who considered a inhomogeneous background field to
evaluate the effective action with the ones provided by the LDE in the same
context so that it could generate the effects of kink configurations beyond
the large $N$ limit considered in Ref. \cite {gnpolymers}.

Another purpose of the work done here is to show the advantages and
reliability of an alternative non-perturbative method like the LDE in the
understanding of the phase diagram of the massless GN model when considered
beyond the large-$N$ limit.  In this case, the massless GN model provides an
excellent testing framework for the following reasons.  As mentioned in the
previous paragraph, large-$N$ results for the critical temperature ($T_c$)
\cite{wrongtc}, critical chemical potential ($\mu_c$) \cite{muc}, and
tricritical points \cite {italianos} are well known \cite{others}.  At the
same time, Landau's theorem \cite{landau,ma} for phase transitions states that
they cannot occur in one space dimension, so that rigorously $T_c=0$, meaning
that the large-$N$ approximation behaves poorly in this case.  These two
extreme results allow us to gauge the LDE performance in connection with the
problem since we know that for $N \to \infty$ our results should converge to
the ``exact", although wrong, large-$N$ result.  {}For finite $N$, on the
other hand, our results should predict smaller values for the critical
temperatures, in accordance with Landau's theorem. At zero temperature and
density, the LDE has been applied to this model with some success
\cite{ldegn,jldamien}, since in this simpler case the LDE could even be summed
to all perturbative orders (at least in the $1/N$ approximation), so that the large order behavior
of the LDE could be investigated. In fact, its convergence properties have
been proved for a particular perturbation series in this context
\cite{jldamien}. At finite temperature, an early application to the GN model
\cite{potef} showed the potentiality of this method. However, the
renormalization program has not been addressed in Ref. \cite{potef}. Here, our
aim is to use all the latest LDE improvements to evaluate the GN effective
potential at finite temperature and density for any value of $N$.
 
This work is organized as follows. In the next section we present the model.
In Sec. III we review the well established $1/N$ results to leading order
considering the following four situations: (a) $T=\mu=0$, (b) $T \ne 0,
\;\mu=0$, (c) $T=0 \;, \mu \ne 0$ and (d) $T \ne 0 \; , \mu \ne 0$.  In Sec.
IV we present the LDE method and the interpolated GN model, evaluating the
effective potential for $N \to \infty$. We show, in accordance with Ref.
\cite{npb}, that when correctly applied, the LDE exactly reproduces, already
at first order, the large-$N$ results. The situation is unchanged, at any
order in $\delta$, provided that one stays within the $N \to \infty$ limit.
This nice result is valid for any parameter values.  In the same section, we
explicitly evaluate the $1/N$ correction that also appears at the first LDE
nontrivial order. In Sec. V the LDE order-$\delta$ results are presented and
compared to the large-$N$ results for the four situations described above. Our
major result is the production of analytical relations for the fermionic mass,
critical temperature, critical chemical potential as well as tricritical
points containing a finite $N$ correction. We show that all these quantities
depend on an optimized mass scale set by the LDE. All the analytical
expressions have been cross checked numerically.  In Sec. VI we contrast the
LDE and the $1/N$ approximation results to leading order and to the next to
the leading order. Our conclusions are presented in Sec. VII. Two appendices
are included to show some technical details and the renormalization for the
interpolated model.

\section{The Gross-Neveu Model}

The Gross-Neveu model is described by the Lagrangian density for a fermion
field $\psi_k$ ($k=1,\ldots,N$) given by \cite{gn}
 
\begin{equation} 
{\cal L} = 
\bar{\psi}_{k} \left( i \not\!\partial\right) \psi_{k} + 
m_F {\bar \psi_k} \psi_k 
+ \frac {g^2}{2} ({\bar \psi_k} \psi_k)^2\;, 
\label{GN} 
\end{equation} 
where the summation over flavors is implicit in the above equation, with e.g.
$\bar{\psi}_k \psi_k = \sum_{k=1}^N \bar{\psi}_k \psi_k$. Since we restrict to
two-dimensional space-time dimension, $\psi_k$ represents a two-component
Dirac spinor for each value of the flavor index $k$. When $m_F=0$ the theory
is invariant under the discrete transformation
 
\begin{equation} 
\psi \to \gamma_5 \psi \,\,\,, 
\end{equation} 
displaying a discrete chiral symmetry (CS). In addition, Eq. (\ref{GN}) has a
global $SU(N)$ flavor symmetry.
 
{}For the studies of the model Eq.  (\ref{GN}) in the large-$N$ limit it is
convenient to define the four-fermion interaction as $g^2 N = \lambda$. Since
$g^2$ vanishes like $1/N$ we study the theory in the large-$N$ limit with
fixed $\lambda$ (see e.g. \cite{coleman}).
 
At finite temperature and density, we can study the model Eq. (\ref{GN}) in
terms of the grand partition function given by
 
\begin{equation} 
Z(\beta,\mu) = {\rm Tr} \exp\left[ - \beta \left( 
H - \mu Q \right) \right] \;, 
\label{Zbetamu} 
\end{equation} 
where $\beta$ is the inverse of the temperature, $\mu$ is the chemical
potential, $H$ is the Hamiltonian corresponding to Eq. (\ref{GN}) and $Q=\int
dx \bar{\psi}_k \gamma_0 \psi_k$ is the conserved charge. Transforming Eq.
(\ref{Zbetamu}) to the form of a path integral in the imaginary-time
(Euclidean) formalism of finite temperature field theory \cite{kapusta}, we
then have

\begin{equation} 
Z(\beta,\mu) = \int \Pi_{k=1}^N D \bar{\psi}_k D \psi_k 
\exp\left\{ - S_{E}[\bar{\psi}_k,\psi_k] \right\} \;, 
\label{func} 
\end{equation} 
where the Euclidean action reads
 
\begin{equation} 
S_{E}[\bar{\psi}_k,\psi_k] = 
\int_0^\beta d \tau \int dx \left[ \bar{\psi}_k 
\left(\not\!\partial + \mu \gamma_0- m_F \right ) \psi_{k} - 
\frac {\lambda}{2N} ({\bar \psi_k} \psi_k)^2 \right] \;,
\label{action} 
\end{equation} 
and the functional integration in Eq. (\ref{func}) is performed over the
fermion fields satisfying the anti-periodic boundary condition in Euclidean
time: $\psi_k(x,\tau) = - \psi_k (x,\tau + \beta)$.

\section{Review of basic $1/N$ results to leading order} 
\label{basic}

Let us briefly review some of the standard large-$N$ results for the GN model.
Considering $m_F=0$ in Eq. (\ref {GN}) we start by looking for a fermionic
mass that can be generated dynamically via radiative corrections. This
exercise will also allow us to set up notation, conventions as well as
reviewing useful formulae to be used within the formalism of finite
temperature and finite density.  As usual, it is useful to rewrite Eq.
(\ref{GN}) expressing it in terms of an auxiliary (composite) field $\sigma$,
so that \cite {coleman}
 
\begin{equation} 
{\cal L} = 
\bar{\psi}_{k} \left( i \not\!\partial\right) \psi_{k} 
-\sigma {\bar \psi_k} \psi_k 
- \frac {\sigma^2 N}{2 \lambda}\;. 
\label{Lsigma} 
\end{equation} 
By using the solution of the equation of motion for $\sigma$ in Eq.
(\ref{Lsigma}) we recover the original model.  {}For renormalization purposes,
we can also add to Eq. (\ref{Lsigma}) a counterterm Lagrangian density, ${\cal
  L}_{ct}$, whose most general form can be expressed as
 
\begin{equation} 
{\cal L}_{ct} = 
\bar{\psi}_{k} \left( i A\not\!\partial\right) \psi_{k} 
-B \sigma {\bar \psi_k} \psi_k 
- C \frac {\sigma^2 N}{2 \lambda} + D {\bar \psi_k} \psi_k + E\sigma + X \;,
\label{lcterm} 
\end{equation} 
where $A,B,C,D,E$ and $X$ are renormalization counterterms with the latter
representing the zero point energy subtraction.

The appearance of a non-vanishing vacuum expectation value for $\sigma$,
$\langle \sigma \rangle = {\bar \sigma_c \neq 0}$ can be associated to a mass
term for the fermion field. This is better studied in terms of the effective
potential for $\sigma$, $V_{\rm eff}(\sigma_c)$. As it is well known, using
the $1/N$ approximation, the large-$N$ expression for the effective potential
is \cite {coleman}
 
\begin{equation} 
\frac {V_{\rm eff}^N}{N}(\sigma_c) =  \frac {\sigma_c^2}{2 \lambda} +
i \int \frac {d^2 p}{(2\pi)^2}
\ln \left(p^2 - \sigma_c^2\right)\;. 
\label{VN} 
\end{equation} 
The above equation can be extended to finite temperature $T$ and density $\mu$
using the usual associations and replacements (see Appendix A for details and
notation), with the result

\begin{eqnarray} 
\frac{V_{\rm eff}^N}{N}(\sigma_c,T,\mu) &=&  \frac {\sigma_c^2}{2 \lambda} -
 \int_p \left[
\omega_{p}(\sigma_c) + T \ln\left( 
1+ \exp\left\{-\left[\omega_{p}(\sigma_c)+\mu\right]/T\right\} 
\right) \right. \nonumber \\ 
&&~~~~~~~~~~~~~~~~~~~~~~~~~~+ \left.  
T \ln\left( 1+\exp\left\{-\left[\omega_{p}(\sigma_c)- 
\mu\right]/T\right\}  
\right) \right] \;,
\label{Vefffull} 
\end{eqnarray} 
where $\omega^2_{p}(\sigma_c) = {\bf p}^2 + \sigma_c^2$ and $\int_p$ denotes
integration over space momentum.  The $T=0,\; \mu=0$ term in Eq.
(\ref{Vefffull}) gives

\begin{eqnarray} 
\frac {V_{\rm eff}^N}{N} (\sigma_c,T=0,\mu=0) &=& 
\frac {\sigma_c^2}{2 \lambda} -
 \int_p \omega_{ p}(\sigma_c) \nonumber \\ 
&=&  \frac {\sigma_c^2}{2 \lambda}  
+\sigma_c^2 
 \left(\frac{e^{\gamma_E} M^2}{\sigma_c^2}\right)^{\epsilon/2}  
\frac{\Gamma\left( \frac{\epsilon}{2} - 1 \right)}{(4 \pi)} 
\nonumber \\ 
&=&  \frac {\sigma_c^2}{2 \lambda} - 
\frac{1}{(4\pi)} \sigma_c^2 
\left [ \frac {2}{\epsilon} + 1 + 
\ln \left ( \frac {M^2}{\sigma_c^2} \right ) + {\cal O}(\epsilon) \right ] \;,
\label{V00} 
\end{eqnarray} 
where $M$ is the arbitrary mass scale introduced by dimensional
regularization.  The divergent term in Eq. (\ref{V00}) can either be rendered
finite, by imposing a renormalization condition directly on the effective
potential by defining a renormalized coupling constant $\lambda_R$, via
 
\begin{equation} 
\frac{d^2 V_{\rm eff}^N}{d \sigma_c^2} = \frac{N}{\lambda_R}\;,
\end{equation} 
or by direct use of the counterterms in Eq. (\ref{lcterm}), in which case we
only require a mass counterterm for the $\sigma$ field that exactly cancels
the divergent term in Eq. (\ref{V00}).  The two renormalizations are of course
equivalent, only differing by a different choice of mass scale in (\ref{V00}).
Choosing the latter form of renormalization, we immediately get the
renormalized relation

\begin{equation} 
\frac{V_{\rm eff}^N}{N}(\sigma_c,T=0,\mu=0) =  \frac {\sigma_c^2}{2 \lambda} -
\frac{1}{(4\pi)} \sigma_c^2 
\left [  1 + \ln \left ( \frac {M^2}{\sigma_c^2} \right ) \right ]\;. 
\label{VTmu=0} 
\end{equation} 
{}From the $T,\; \mu$ dependent term of Eq. (\ref{Vefffull}),
 
\begin{equation} 
 \frac{V_{\rm eff}^N}{N}(\sigma_c,T, \mu)
= - \int_p \left[ 
T \ln\left( 
1+ \exp\left\{-\left[\omega_{p}(\sigma_c)+\mu\right]/T\right\} 
\right) + 
T \ln\left( 1+\exp\left\{-\left[\omega_{p}(\sigma_c)-\mu\right]/T\right\}  
\right) \right] \;, 
\label{VTmu} 
\end{equation} 
we can take the limit $T\to 0$ and perform the momentum integral to obtain the
finite result
 
\begin{equation} 
 \frac{V_{\rm eff}^N}{N}(\sigma_c,T=0, \mu\neq 0)
=  \frac{1}{2 \pi} \theta(\mu - \sigma_c)
\left[ \sigma_c^2 \ln \left( \frac{\mu + \sqrt{\mu^2 - 
\sigma_c^2}}{\sigma_c} 
\right) - \mu \sqrt{\mu^2 - \sigma_c^2} \right]\;. 
\label{VT0mu} 
\end{equation} 
The general expression Eq. (\ref{VTmu}) for $T \neq 0$ and $\mu \neq 0$,
however, does not have a close analytical expression, but we can express it in
terms of a high temperature expansion, in powers of $\mu/T$ and $\sigma_c/T$,
in analogous form as it is done for the bosonic like temperature dependent
momentum integrals \cite{weldon}.  Defining the function $I_1(a,b)$ by
 
\begin{equation} 
I_1(a,b) = \int_0^\infty dx  \left[ \ln \left( 1+e^{-\sqrt{x^2+a^2}-b} \right) 
+ \ln \left( 1+e^{-\sqrt{x^2+a^2}+b} \right) \right]\;, 
\label{Jab} 
\end{equation} 
Eq. (\ref{VTmu}) can be written as
 
\begin{equation} 
\frac{V_{\rm eff}^N}{N}(\sigma_c,T, \mu)
= -  T^2 \frac{\sqrt{2}}{(2 \pi)^{\frac{1}{2}} \Gamma\left( 
\frac{1}{2} \right)} I_1 \left(\sigma_c/T,\mu/T \right)\;. 
\label{VJ} 
\end{equation} 
We now take $\sigma_c/T = a \ll 1$ and $\mu/T = b \ll 1$.  Expanding Eq.
(\ref{Jab}) in powers of $a$ and $b$, the result is finite and given by
\cite{zhou}
 
\begin{equation}
I_1 (a\ll 1,b\ll 1) = \frac{\pi^2}{6} + \frac{b^2}{2} - \frac{a^2}{2}
\ln \left(\frac{\pi}{a} \right) - \frac{a^2}{4}(1-2\gamma_E)
- \frac{7 \zeta(3)}{8 \pi^2} a^2 \left(b^2 + \frac{a^2}{4} \right)
+{\cal O}(a^2 b^4, a^4 b^2)\;, 
\label{JT} 
\end{equation}
where $\zeta(3) \simeq 1.202$.  Using Eq. (\ref{JT}) in Eq. (\ref{VJ}), the
high temperature expansion for $ V_{\rm eff}(\sigma_c,T, \mu)$ becomes

\begin{equation}
\frac{V_{\rm eff}^N}{N}(\sigma_c,T \neq 0, \mu\neq 0)
=   \frac{\sigma_c^2}{2 \pi} \ln\left(\frac{\pi T}{\sigma_c} \right)
-  \frac{\sigma_c^2 \gamma_E}{2 \pi} +  \frac{\sigma_c^2}{4 \pi}
+ \frac{7 \zeta(3)}{8 \pi^3 T^2} \sigma_c^2 \left(\mu^2 +
\frac{\sigma_c^2}{4} \right)  
+ {\cal O} \left( \frac{\sigma_c^2 \mu^4}{T^4},\frac{\sigma_c^4\mu^2}{T^4}   
\right)\;, 
\label{VhighT} 
\end{equation}
where we have dropped terms that do not depend on $\sigma_c$. Let us now
review four important situations

\subsection{The $T=\mu=0$ case: the fermionic mass at large-$N$}

At $T=\mu=0$ the large-$N$ effective potential is given by Eq. (\ref{VTmu=0}),
 
\begin{equation} 
\frac{V_{\rm eff}^{N}}{N}(\sigma_c,T=0,\mu=0) =  
\frac {\sigma_c^2}{2 \lambda} -
\frac{1}{4\pi} \sigma_c^2 
\left [  1 + \ln \left ( \frac {M^2}{\sigma_c^2} \right ) \right ]\;. 
\end{equation} 
We can already check here that, contrary to the classical (tree-level)
potential where the minimum occurs at ${\bar \sigma_c}=0$, one now has
symmetry breaking that is quantum generated since the minimum of the effective
potential occurs at a non-vanishing value given by
 
\begin{equation} 
\frac{\partial V_{\rm eff}^N(\sigma_c,T=0,\mu=0)}{\partial \sigma_c} 
\Bigr|_{\sigma_c = \bar{\sigma}_c } =0\;, 
\end{equation} 
where ${\bar \sigma_c}$ sets the large-$N$ result for the fermionic mass, at
$T=\mu=0$, as
 
\begin{equation} 
m_F(0)= {\bar \sigma}_c=M \exp\left(-\frac{\pi}{\lambda}\right).
\label{mF} 
\end{equation}

\subsection{The $T \ne 0$ and $\mu = 0$ case: the critical temperature $T_c$}

Physically, the case $\mu = 0$ means that the number of fermions and
anti-fermions are the same.  {}For this case, from Eq. (\ref{VTmu=0}) and
using the result (\ref{VhighT}) for the finite temperature term for the
effective potential, we have that
 
\begin{equation} 
\frac{V_{\rm eff}^{N}}{N}(\sigma_c,T\neq 0,\mu=0) =  
\frac {\sigma_c^2}{2 \lambda} -
\frac{1}{2\pi} \sigma_c^2 
\ln \left ( \frac {M}{\pi T} \right ) 
-  \frac{\sigma_c^2 \gamma_E}{2 \pi} 
+ \frac{7 \zeta(3)}{32 \pi^3 T^2} \sigma_c^4 
\label{vtN} 
\;, 
\end{equation} 
which shows chiral symmetry restoration at a critical temperature $T_c$ given
by
 
\begin{equation} 
T_c = m_F(0) \frac{e^{\gamma_E}}{\pi} \simeq 
0.567 \; m_F (0), 
\label{Tc} 
\end{equation} 
in accordance with Ref. \cite {wrongtc}.  This transition is second order, as
can be easily checked by writing Eq. (\ref{vtN}) in the form
 
\begin{equation} 
\frac{V_{\rm eff}^{N}}{N}(\sigma_c,T\neq 0,\mu=0) = \frac {1}{2 \pi}
\ln\left(\frac{T}{T_c} \right) \sigma_c^2 
+ \frac{7 \zeta(3)}{32 \pi^3 T^2} \sigma_c^4  
\label{vtc} 
\;, 
\end{equation} 
which for $T$ close to $T_c$ gives a temperature dependent vacuum expectation
for $\sigma_c(T) \equiv m_F(T)$ in the form
 
\begin{equation} 
m_F(T) \simeq \sqrt{\frac{8 \pi^2}{7 \zeta(3)}} T_c  
\sqrt{\frac{T_c-T}{T_c}}\;, 
\end{equation} 
that shows a continuous transition at the critical point $T=T_c$.  {}For
$T\geq T_c$ the fermion mass $m_F(T)$ vanishes, restoring the chiral symmetry.

\subsection{The $T=0$ and $\mu \ne 0$ case: the critical density $\mu_c$} 
\label{T0mufinite} 
 
{}For the case of zero temperature but unequal number of fermions and
anti-fermions, $\mu \neq 0$, one considers Eqs. (\ref{VTmu=0}) and
(\ref{VT0mu}), which give the expression for the effective potential at finite
density but zero temperature. Then,

\begin{equation} 
\frac{V_{\rm eff}^{N}}{N}(\sigma_c,T=0,\mu \neq 0) =  
\frac {\sigma_c^2}{2 \lambda} -
\frac{1}{4\pi} \sigma_c^2 
\left [  1 + \ln \left ( \frac {M^2}{\sigma_c^2} \right ) \right ] 
+ \frac{1}{2 \pi} \theta(\mu - \sigma_c) 
\left[ \sigma_c^2 \ln \left( \frac{\mu + \sqrt{\mu^2 - \sigma_c^2}}{\sigma_c} 
\right) - \mu \sqrt{\mu^2 - \sigma_c^2} \right]\;. 
\label{VT0mu2} 
\end{equation} 
{}For $\mu \leq \sigma_c$ the minimum of the potential reproduces the same
result as Eq. (\ref{mF}).  {}For $\mu > \sigma_c$ we have from Eq.
(\ref{VT0mu2}) that
 
\begin{equation} 
\bar{\sigma}_c (\mu) = \sqrt{m_F(0) \left[ 2 \mu - m_F(0) \right]}\;, 
\end{equation} 
which is valid for $m_F(0)/2 \leq \mu <m_F(0)$.  In the case ${\bar \sigma}_c
= m_F(0)$ and from Eq. (\ref{VT0mu2}) we obtain
 
\begin{equation} 
\frac{V_{\rm eff}^{N}}{N}(\sigma_c,T=0,\mu < \sigma_c) = 
-\frac{1}{4 \pi}m_F^2(0)\;.
\end{equation} 
{}For $\mu < m_F(0)/2$, $V_{\rm eff}^{N}(\sigma_c,T=0,\mu \neq 0)$ will have a
minimum for ${\bar \sigma}_c=m_F(0)$ and a maximum for ${\bar \sigma}_c=0$.
{}For $m_F(0)/2 \leq \mu <m_F(0)$, $V_{\rm eff}^{N}(\sigma_c,T=0,\mu \neq 0)$
has minima at ${\bar \sigma}_c=0$ and ${\bar \sigma}_c=m_F(0)$ and a maximum
at $\bar{\sigma}_c(\mu)$. There is a value of $\mu$ where both minima satisfy
 
\begin{equation} 
V_{\rm eff}^{N}({\bar \sigma}_c=m_F(0),T=0,\mu_c) =
V_{\rm eff}^{N}({\bar \sigma}_c=0,T=0,\mu_c)\;,
\label{V=V} 
\end{equation} 
where

\begin{equation} 
\mu_c =\frac{ m_F(0)}{\sqrt{2}} \,\,, 
\label{largeNmuc} 
\end{equation} 
in agreement with Refs. \cite{muc,others}.  This result indicates a first
order phase transition: for values of $\mu > \mu_c$, ${\bar \sigma}_c=m_F(0)$
is a local minimum (false vacuum), while ${\bar \sigma}_c=0$ is a global
minimum (true vacuum). {}For $\mu < \mu_c$ the two vacua get interchanged.

\subsection{The $T \ne 0$ and $\mu \ne 0$ case} 
\label{tricritsec} 
 
This situation can be analyzed numerically only.  On the ($T,\mu$) plane one
can identify a tricritical point $P_{tc}$ where the first and second order
transition lines meet. In units of $m_F(0)$, the numerical result is
$P_{tc}=(T_{tc},\mu_{tc})=(0.318, 0.608)$ \cite{italianos}. The large-$N$
phase diagram will be presented in subsection V.D together with the LDE
result.

\section{The effective potential for the interpolated theory}

Let us now turn our attention to the implementation of the LDE procedure
within the GN model.  According to the usual LDE interpolation prescription
\cite {linear} (for a long, but far from complete list of references on the
method, see \cite{early}) the deformed {\it original} four fermion theory
displaying CS reads

\begin{equation}
{\cal L}_{\delta}(\psi, {\bar \psi}) =
\bar{\psi}_{k} \left( i \not\!\partial\right) \psi_{k} +
\eta (1-\delta) {\bar \psi_k} \psi_k
+ \delta \frac {\lambda}{2N} ({\bar \psi_k} \psi_k)^2\;.
\label{GNlde}
\end{equation}

\noindent
So, that at $\delta=0$ we have a theory of free fermions.  Now, the
introduction of an auxiliary scalar field $\sigma$ can be achieved by adding
the quadratic term,

\begin{equation} 
- \frac{ \delta N}{2 \lambda} \left ( \sigma +  
\frac {\lambda}{N} {\bar \psi_k} \psi_k \right )^2 \,, 
\end{equation} 
to ${\cal L}_{\delta}(\psi, {\bar \psi})$. We are then lead to the
interpolated model
 
\begin{equation} 
{\cal L}_{\delta} = 
\bar{\psi}_{k} \left( i \not\!\partial\right) \psi_{k} - 
\delta \sigma {\bar \psi_k} \psi_k - \eta (1-\delta) {\bar \psi_k} \psi_k 
- \frac {\delta N }{2 \lambda } \sigma^2 + {\cal L}_{ct,\delta}  \;, 
\label{GNdelta} 
\end{equation} 
 
\noindent 
where the counterterm Lagrangian density, ${\cal L}_{ct,\delta}$, has the same
polynomial form as in the original theory, Eq. (\ref{lcterm}), while the
coefficients are allowed to be $\delta$ and $\eta$ dependent
\cite{ldegn,prd1}.  Details of the renormalization process for the
interpolated model are given in the appendix.  Note that the same
interpolation of the form (\ref{GNdelta}) was also used in Ref.  \cite{ijmpe},
but it is different from the ones used in Refs.
\cite{ldegn,npb,potef,klimenko}. In those references the interpolation was not
carried out in the original four fermion theory but on its bosonized version.
However, we argue that the present choice is more adequate because at
$\delta=0$ one has only free fermions. Otherwise, the quadratic bosonic term
$\sigma^2/(2\lambda)$ survives at $\delta=0$ and the theory looks to be
composed by free fermions and bosons but this is misleading since, by the
equations of motion, $\sigma = (\lambda/N) {\bar \psi}_k \psi_k$.  {}From the
Lagrangian density in the interpolated form, Eq.  (\ref{GNdelta}), we can
immediately read the corresponding new Feynman rules in Minkowski space. Each
Yukawa vertex carries a factor $-i \delta$ while the (free) $\sigma$
propagator is now $-i \lambda/(N \delta)$. The LDE dressed fermion propagator
is

\begin{equation} 
S_F(p)=\frac{i}{\not \! p - \eta_*+i\epsilon}\;, 
\label{SF} 
\end{equation} 
where $\eta_*= \eta -(\eta - \sigma_c)\delta$.

\subsection{The LDE Effective Potential in the Large-$N$ Limit} 
\label{subseclargeNlde}

In the following we start our analysis of the LDE effective potential in the
large-$N$ limit, {\it i.e.}, the LDE is applied directly to the effective
potential {\it after} the large-$N$ limit was taken. In practice, as already
emphasized, within the LDE one perturbs in powers of $\delta$ only, without
having to notice the actual powers of $N$ as we do here. However, in this
subsection, we do the $\delta$ expansion within the $1/N$ leading order only
to be able to compare both methods. In other words, we first want to establish
the reliability of the LDE-PMS by considering $N \to \infty$ in order to
further compare our results with the ``exact" ones furnished by large-$N$
approximation. The reader should not be confused by this apparent mixed type
of expansion. Using the new {}Feynman rules from the interpolated model
(\ref{GNdelta}), in the large-$N$ limit, one can write the effective
potential, analogous \footnote{In this subsection we shall work in an
  arbitrary number of space-time dimensions to emphasize the generality of our
  optimization procedure.} to Eq. (\ref{VN}), to {\it arbitrary} orders in
$\delta$ as \cite{npb,potef,ijmpe}
 
\begin{equation} 
\frac {V_{{\rm eff}}}{N}(\sigma_c, \eta_*,N \to \infty)= \delta
\frac {\sigma_c^2}{2 \lambda} + 
i \int \frac {d^d p}{(2\pi)^d} {\rm tr}\ln \left(\not \! p - \eta_* 
\right)\;\;,
\label{veffNlde} 
\end{equation} 
 
\noindent 
which can be seen as obtained directly from the series containing all
perturbative terms, Eq. (\ref {VN}), simply by performing the replacement
$\sigma_c \to \eta_*$. Developing Eq. (\ref{veffNlde}) to first order in
$\delta$ one obtains
 
\begin{equation}
\frac{V_{{\rm eff},\delta^1}}{N} (\sigma_c,\eta,N\to \infty)= \delta
\frac {\sigma_c^2}{2 \lambda} +
i  \int \frac {d^d p}{(2\pi)^d} {\rm tr}\ln \left(\not \! p - \eta \right)+
\delta  i\int \frac {d^d p}{(2\pi)^d} {\rm tr}
\frac {\eta-\sigma_c}{\not \! p - \eta + i \epsilon} \;\;.
\label{vlde2}
\end{equation}

{}Fixing $\delta=1$, we now optimize $V_{\rm eff}$ for $\eta$ as in most of
the previous references on the LDE method by using the principle of minimal
sensitivity (PMS). In the PMS procedure one requires that a physical quantity
$\Phi^{(k)}$, that is calculated perturbatively to some $k$-th order in
$\delta$, be evaluated at the point where it is less sensitive to this
parameter.  This criterion then translates into the variational relation
\cite{pms}

\begin{equation} 
\frac {d \Phi^{(k)}}{d \eta}\Big |_{\bar \eta, \delta=1} = 0 \;. 
\label{PMS} 
\end{equation} 

The optimum value $\bar \eta$ that satisfies Eq. (\ref{PMS}) must be a
function of the original parameters, including the couplings, thus generating
non-perturbative results.  In the following we will apply Eq. (\ref{PMS})
directly to the effective potential.  Applying the PMS to Eq. (\ref{vlde2})
immediately gives the result ${\bar \eta} = \sigma_c$, which then recovers
{\it exactly} the large-$N$ result, Eq. (\ref{VN}). This same trend holds at
any temperature and/or density, value of the coupling $\lambda$, as well as
number of space-time dimensions since there is no need to perform the
integrals explicitly. Also, as shown in Ref. \cite{npb}, the inclusion of
higher order terms does not spoil this nice result, provided we stay within
the large-$N$ limit, since they are all of the form $\delta^k f_k ( \eta -
\sigma_c)^k$, where $k \ge 2$. Note that the PMS admits another solution given
by

\begin{equation} 
\frac {d}{d \eta} \left [\int \frac {d^d p}{(2\pi)^d}  {\rm tr}  
\frac{i}{\not \! p - \eta+i \epsilon} \right ]_{\bar \eta} = 0 \;.
\label{other} 
\end{equation} 
However, this solution, which depends only on scales introduced by the
regularization process, and thus is not proportional to the basic scale of the
model after dimensional transmutation: $M e^{-\pi/\lambda}$, can be taken as
unphysical \cite{npb} since it brings no information about the theory being
studied.

\begin{figure}[htb]
  \vspace{0.5cm}
  \epsfig{figure=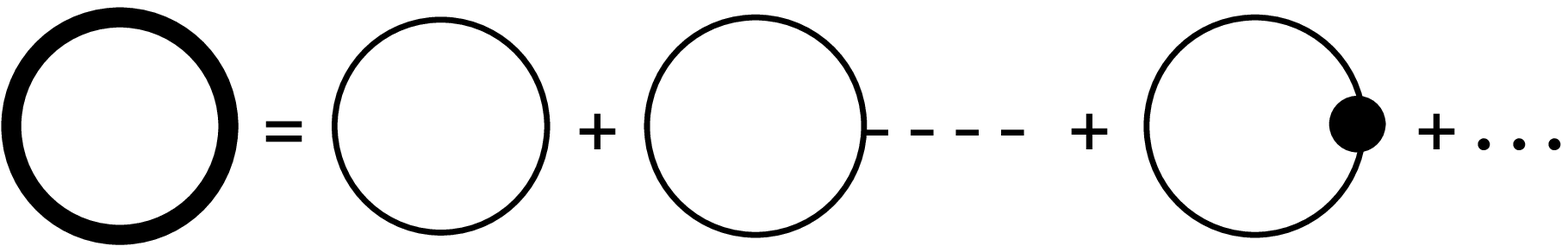,angle=0,width=12cm}
\caption[]{\label{gnfig0} Diagrammatic relation between  
  Eqs. (\ref{veffNlde}) and (\ref{vlde2}). The thick continuous line on the
  LHS represents the LDE dressed fermionic propagator which is $\eta_*$
  dependent. The first diagram on the RHS is an order-$\delta^0$ vacuum graph.
  The dashed line represents the auxiliary field, $\sigma_c$, while the black
  dot stands for the $\delta \eta$ vertex. The last two graphs on the RHS are
  of order-$\delta$.}
\end{figure}

\noindent
One can spot a subtle point associated to the LDE evaluation of the effective
potential by looking at the diagrams considered to order-$\delta$ in Eq. (\ref
{vlde2}). These graphs are displayed in Fig. \ref {gnfig0}, which shows the
diagrammatic relation between Eqs. (\ref{veffNlde}) and (\ref{vlde2}). Note
that Eq. (\ref {vlde2}) contains two $\sigma_c$ independent (vacuum) terms
that would be neglected in most evaluations since they are irrelevant as far
as CS breaking/restoration are concerned. However, as pointed out in Ref.
\cite{npb}, they are $\eta$ dependent an so must be considered until the
theory is optimized. One can also easily see that, in fact, those
contributions are responsible for the quick LDE convergence towards the exact
large-$N$ result already at the first non-trivial order. The explicit form of
the LDE large-$N$ effective potential is quickly obtained by applying the
finite $T$ and $\mu$ rules described in the previous section to Eq. (\ref
{veffNlde}). In particular, from Eqs.  (\ref{VTmu=0}) and (\ref{VJ}) for
$d=2$, one then obtains the renormalized result

\begin{equation} 
\frac{V_{{\rm eff},\delta^1}}{N} (\sigma_c,\eta_*,N\to \infty)=
\delta  \frac {\sigma_c^2}{2 \lambda} -
\frac{1}{2\pi} \left \{ \eta_*^2 \left [ \frac {1}{2} +
\ln \left( \frac {M}{\eta_*} \right ) \right] +  
2 T^2 I_1\left( \frac{\eta_*}{T},\frac{\mu}{T}\right) \right \} \;. 
\label{VJdelta} 
\end{equation} 
 
Expanding Eq. (\ref{VJdelta}) to order $\delta$ at $T=\mu=0$ one retrieves, as
before, the PMS result ${\bar \eta} = \sigma_c$ that reproduces the large-$N$
result. The same optimum value for $\eta$ appears when we express Eq.
(\ref{VJdelta}) in the other limits studied in the previous section. At the
same time, as discussed above, the PMS gives another solution, Eq.
(\ref{other}), which is ${\bar \eta} = M/e$. Rather curiously, this result
coincides with ${\bar \sigma_c}=m_F(0)$ evaluated at $\lambda= \pi$.

\subsection{The Effective Potential to Order $\delta$: going  
  beyond the large-$N$ limit}

\begin{figure}[htb]
  \vspace{0.5cm}
  \epsfig{figure=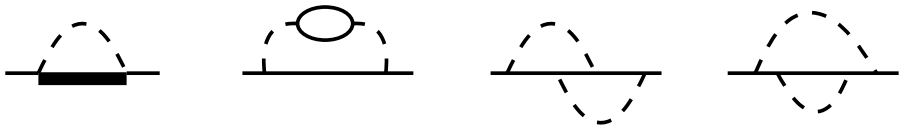,angle=0,width=12cm}
\caption[]{\label{gnfig1} Contributions to the fermion self-energy
  up to order $\delta^2$. The continuous thick line represents the LDE,
  $\eta_*$ dependent, dressed fermion propagator while the thin lines
  represent the $\eta$ dependent propagators. The dashed line represents the
  scalar auxiliary field, $\sigma$. Tadpole diagrams are not shown since they
  do not contribute to $V_{\rm eff}$}. The second graphs represents a
correction to the $\sigma$ propagator while the third has corrections to the
Yukawa vertex. The first and fourth are exchange (rainbow) type of graphs.
\end{figure}

\begin{figure}[htb]
  \vspace{0.5cm}
  \epsfig{figure=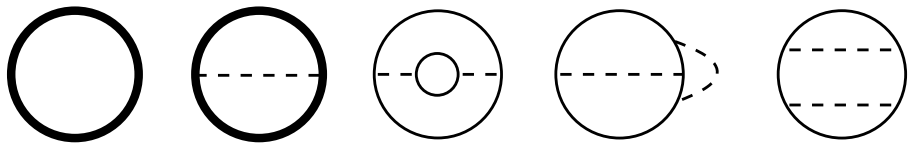,angle=0,width=12cm}
\caption[]{\label{gnfig2}
  Feynman diagrams contributing to the effective potential at order
  $\delta^2$. Note that the first and second have a $1/N$ dependence while the
  last two have a $1/N^2$ dependence.  The thick lines in the first two graphs
  represent the LDE, $\eta_*$, propagators which, as discussed in the text
  must be further expanded to order-$\delta^2$.}
\end{figure}

In the following we show how the optimization procedure implemented by the LDE
improves over the large-$N$ results. Here we revert to the usual LDE procedure
by expanding the effective potential in powers of $\delta$ only.  This
quantity can be expressed in terms of the full fermionic self-energy, whose
terms contributing up to order $\delta^2$ are displayed in {}Fig.
\ref{gnfig1}.  The corresponding contributions to the effective potential up
to order $\delta^2$ are shown in {}Fig. \ref{gnfig2}. This nicely illustrates
how the LDE incorporates, at the same perturbative order, graphs that have
different $N$ dependence.

Let us now consider the second graph of Fig. \ref {gnfig2} which, for
comparison purposes, has been neglected in the previous subsection. It reads

\begin{equation}
 \frac{V_{{\rm eff},\delta^1}^{(a)}}{N} (\eta_*)=
-i \frac {1}{2} \int \frac {d^2 p}{(2\pi)^2} 
{\rm tr} \left [\frac {\Sigma_a(\eta_*)}{\not \! p - \eta_* + 
i \epsilon} \right ]\;,
\label{VN1} 
\end{equation} 
where the trace is over Dirac's matrices only \footnote {The factor $-1$
  corresponding to a closed fermionic has already been taken into account
  \cite {root}.}. The term $\Sigma_a$ represents the first contribution of
{}Fig. \ref{gnfig1} to the fermion self-energy,
 
\begin{equation} 
\Sigma_a (\eta_*) = -\delta  \left (\frac {\lambda}{N} \right ) 
i\int \frac {d^2 q}{(2 \pi)^2} \frac {1}{\not \! q - \eta_*+i \epsilon}\;. 
\label{Sigma1} 
\end{equation} 
 
Originally, Root \cite {root} has summed up all $1/N$ corrections to next to
the leading order and by further perturbatively expanding his result one
easily retrieves Eq. (\ref {VN1}).

Then, using the rules described previously, at finite temperature and density,
one finds, after taking the trace and renormalizing with the fermionic mass
counterterm (see the appendix), the result for Eq. (\ref{VN1})

\begin{equation} 
 \frac{V_{{\rm eff},\delta^1}^{(a)}}{N}(\sigma_c,\eta_*,T,\mu) =
\delta \frac {\lambda}{4\pi^2\:N}\; \left\{ \eta^2_*
\left [ \ln \left ( \frac {M}{\eta_*}  \right ) -  
I_2(\eta_*/T,\mu/T) \right ]^2 +T^2 J^2_2(\eta_*/T,\mu/T) \right\}\;, 
\label{DVN1} 
\end{equation} 
where

\begin{eqnarray} 
I_2(a,b) &=& - 2 \frac{\partial I_1(a,b)}{\partial a^2} 
\nonumber \\ 
&=& \int_0^\infty \frac {d x}{\sqrt{x^2 +a^2}} \left( 
\frac{1}{e^{\sqrt{x^2 + a^2} + b}+1}+\frac{1}{e^{\sqrt{x^2 + a^2} - 
b}+1} \right) \;, 
\label{I2ab} 
\end{eqnarray} 
 
\noindent 
with $I_1(a,b)$ defined by Eq. (\ref{Jab}) where, now, $a=\eta_*/T$, and
similarly
 
\begin{eqnarray} 
J_2(a,b)
= \sinh(b) \int_0^\infty d x 
\frac{1}{\cosh(\sqrt{x^2 + a^2})+\cosh(b)}  \;. 
\label{J2ab} 
\end{eqnarray} 

\noindent
The (finite) contributions given by the term proportional to $I_2(a,b)$ and
$J_2(a,b)$ originate from the summation over Matsubara frequencies of the $T$
and $\mu$ dependent contributions, more precisely from Eqs. (\ref{sum2}) and
(\ref{sum3}) respectively (see Appendix A for details).  Note that the
divergence is only contained, at this order, in the $T=0$, $\mu=0$ part, which
is renormalized by standard counterterms in the $\overline{ \rm MS}$ scheme
(see Appendix B for details).
 
{}Finally, by summing the contribution Eq. (\ref{DVN1}) to the effective
potential expression to leading order in $N$, Eq. (\ref{VJdelta}), and
expanding the resulting expression one obtains the complete LDE expression to
order-$\delta$

\begin{eqnarray}
\frac{V_{{\rm eff},\delta^1}}{N} (\sigma_c, \eta, T, \mu) &=&
\delta \frac {\sigma_c^2}{2 \lambda} - 
 \frac{1}{2\pi} \left \{ \eta^2 \left [ \frac {1}{2} + \ln \left ( 
\frac {M}{\eta} \right ) \right ] + 2 T^2 I_1(\eta/T,\mu/T) \right \} 
\nonumber \\ 
&+& \delta \frac{\eta(\eta-\sigma_c)}{\pi} 
\left[\ln\left(\frac{M}{\eta}\right) - I_2 (\eta/T,\mu/T) \right] 
\nonumber \\ 
&+& \delta \frac {\lambda}{4\pi^2\:N} \left \{ \eta^2  
  \left [ \ln \left ( \frac {M}{\eta}  \right ) - I_2(\eta/T,\mu/T)
\right ]^2 + T^2 J^2_2(\eta/T,\mu/T) \right \}\;.
\label{Vdelta1}
\end{eqnarray} 
 
\noindent 
Notice once more, from Eq. (\ref{Vdelta1}), that our first order takes into
account the first next to leading order correction to the large-$N$ result.

When considering the case $T=0, \mu \ne 0$ in Eq. (\ref{Vdelta1}) one can take
the limit $T\to 0$, which then gives for the functions $I_1$, $I_2$ and $J_2$
the results

\begin{eqnarray} 
&& \lim_{T\to 0} T^2 I_1(\eta/T,\mu/T)= 
- \frac{1}{2} \theta(\mu - \eta) 
\left[ \eta^2 \ln \left( \frac{\mu + \sqrt{\mu^2 -  
\eta^2}}{\eta} 
\right) - \mu \sqrt{\mu^2 - \eta^2} \right]\;, 
\label{J2T0} 
\\ 
&&  
\lim_{T \to 0} I_2(\eta/T,\mu/T) =  \theta(\mu-\eta) 
\ln\left(\frac{\mu +\sqrt{\mu^2 - \eta^2}}{\eta} \right)\;,
\label{I2T0} 
\\ 
&&
\lim_{T \to 0} T J_2(\eta/T,\mu/T) =  {\rm sgn}(\mu) \theta(\mu-\eta)
\sqrt{\mu^2 - \eta^2} \;. 
\label{JJ2T0} 
\end{eqnarray} 
 
\noindent 
Note that all the above results vanish for $\mu < \eta$.  When considering the
case $T \ne 0$ and $\mu=0$ the high temperature limit, $\eta/T\ll 1$ and
$\mu/T\ll 1$ will prove to be useful. For $I_1$, this approximation follows
from Eq. (\ref{JT}), while for $I_2$ it can be obtained using Eq.
(\ref{I2ab}) which yields
 
\begin{equation} 
I_2(a,b) = \ln\left(\frac{\pi}{a}\right) - \gamma_E  
+\frac{7 \xi(3)}{4 \pi^2} \left(b^2+\frac{a^2}{2}\right) +  
{\cal O}(a^4,b^4)\;. 
\label{highTI2} 
\end{equation} 
In the case $T \ne 0$ and $\mu \ne 0$ the integrals $I_1$, $I_2$ and $J_2$
will be handled numerically.

\section{Optimization and Numerical Results Beyond Large-$N$}

\begin{figure}[htb]
  \vspace{0.5cm}
  \epsfig{figure=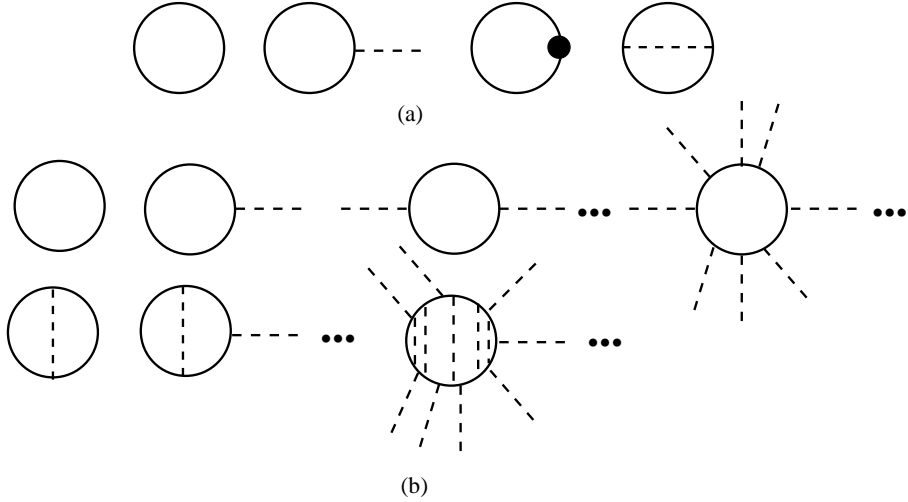,angle=0,width=12cm}
\caption[]{\label{gnfig4}
  (a) Feynman diagrams contributing to the effective potential at order
  $\delta$ prior to optimization. (b) The infinite set of graphs contributing
  to the optimized effective potential.}
\end{figure}

Before proceeding to the specific $d=2$ case, considered in this work, let us
apply the PMS to the most general order-$\delta$ effective potential which is
given by

\begin{equation}
\frac{V_{{\rm eff},\delta^1}}{N} (\sigma_c,\eta)= \delta
\frac {\sigma_c^2}{2 \lambda} +
i  \int \frac {d^d p}{(2\pi)^d} {\rm tr}\ln \left(\not \! p - \eta \right)+
\delta  i\int \frac {d^d p}{(2\pi)^d} {\rm tr}
\frac {\eta-\sigma_c}{\not \! p - \eta + i \epsilon}
+ \frac {\delta \lambda}{2N} {\rm tr}\left [i\int \frac {d^d p}{(2\pi)^d}
\frac {1}{\not \! p - \eta + i \epsilon}\right ]^2 \,\,.
\label{general}
\end{equation}
This exercise will help the reader to visualize the way the LDE-PMS resumms
the perturbative series.  After taking the traces in Eq. (\ref{general}) one
obtains

\begin{eqnarray}
\frac{V_{{\rm eff},\delta^1}}{N} (\sigma_c,\eta)&= &\delta
\frac {\sigma_c^2}{2 \lambda} +
i  \int \frac {d^d p}{(2\pi)^d} \ln \left( p^2 - \eta^2 \right)\delta
+2 i\int \frac {d^d p}{(2\pi)^d}
\frac {\eta(\eta-\sigma_c)}{p^2 - \eta^2 + i \epsilon} \nonumber \\
&+& \delta \frac{\lambda}{N} \eta^2\left [i\int \frac {d^d p}{(2\pi)^d} 
\frac {1}{p^2 - \eta^2 + i \epsilon} \right ]^2 + \delta \frac{\lambda}{N} 
\left [i\int \frac {d^d p}{(2\pi)^d} \frac {p_0}{p^2 - \eta^2 + i \epsilon} 
\right ]^2 \,\,.
\end{eqnarray}
Then, setting $\delta=1$ and applying the PMS one gets

\begin{eqnarray}
0&=&\left \{ \left [ \eta - \sigma_c + \eta \frac {\lambda}{N}
\left (i\int \frac {d^d p}{(2\pi)^d} \frac {1}{(p^2 - \eta^2 + i \epsilon)} 
\right ) \right ] \left( 1 + \eta \frac{d}{d\eta} \right ) 
\left [i\int \frac {d^d p}{(2\pi)^d} \frac {1}{(p^2 - \eta^2 + i \epsilon)}  
\right . \right ]\nonumber \\
 &+&\left . \frac{\lambda}{N} \left (i\int \frac {d^d p}{(2\pi)^d} 
\frac {p_0}{(p^2 - \eta^2 + i \epsilon)}  \right )\frac{d}{d \eta}
\left (i\int \frac {d^d p}{(2\pi)^d} \frac {p_0}{(p^2 - \eta^2 + i \epsilon)}  
\right ) \right \}\Bigr|_{\eta = {\bar \eta}}\,\,.
\label{pmsselfconsistent}
\end{eqnarray}
As one can see in Appendix A (Eq. (\ref {sum3})) the last term of
the equation above only survives when $\mu \ne 0$. In the case $\mu=0$, Eq.
(\ref {pmsselfconsistent}) factorizes in a nice way which allows us to
understand the way the LDE-PMS procedure resums the series producing
non-perturbative results.  With this aim one can easily check that (at
$\delta=1$)

\begin{equation}
\Sigma_a(\mu=0,\eta,T) = - \frac{\lambda}{N} \eta 
\left [i\int \frac {d^d p}{(2\pi)^d} \frac {1}{(p^2 - \eta^2 + i \epsilon)} 
\right ] \;\;.
\end{equation}
Then, when $\mu=0$ the PMS equation factorizes to

\begin{equation}
[ {\bar \eta} - \sigma_c - \Sigma_a(\mu=0,\eta,T)]
\left( 1 + \eta \frac{d}{d\eta} \right ) 
\left [i\int \frac {d^d p}{(2\pi)^d} \frac {1}{(p^2 - \eta^2 + i \epsilon)}  
\right ]\;,
\label{selfconsistent}
\end{equation}
which leads to the self consistent relation

\begin{equation}
{\bar \eta} =\sigma_c + \Sigma_a(\mu=0,\eta,T) \,\,\,,
\label{selfcons}
\end{equation}
which is valid for any temperature and number of space-time dimensions. In
this way the LDE fermionic loops get dressed by $\sigma_c$ as well as rainbow
(exchange) type of self energy terms like the first graph of figure
\ref{gnfig2}. Typical optimized Feynman graphs are shown in {}Fig. \ref
{gnfig4}. The graphs in part (a) of that figure represent the order-$\delta$
contributions prior to optimization while the infinite set shown in part (b)
represents the non perturbative optimized result. Note how the graphs are
dressed by rainbow type of self energy contributions. This was expected since
at order-$\delta$ the perturbative LDE effective potential receives
information about this type of topology only. If one proceeds to
order-$\delta^2$ information about corrections to the scalar propagator as
well as Yukawa vertex (see figure \ref {gnfig1}) will enter the perturbative
effective potential. Then the PMS will dress up these perturbative
contributions and so on. In other words, the simple evaluation of a first
topologically distinct graph will bring non perturbative information
concerning that type of contribution. Figure \ref{gnfig4} clearly shows that
the LDE-PMS re-sums all powers of $1/N$ corresponding to the (rainbow) class of
graph.

Note that the mathematical possibility

\begin{equation}
i\int \frac {d^d p}{(2\pi)^d} \frac {1}{(p^2 - 
{\bar \eta}^2 + i \epsilon)} =0 \,\,,
\end{equation}
corresponds to the unphysical solution discussed in Sec. IVA (Eq.
(\ref{other})).

Having illustrated the way the LDE-PMS resummation works, let us concentrate
in the $d=2$ case by collecting our results for the complete order-$\delta$
effective potential, Eq. (\ref {Vdelta1}), once the optimization equation is
applied to it.  {}Using the PMS procedure we then obtain, from Eq.
(\ref{Vdelta1}) at $\eta = {\bar \eta}$, the general result
 
\begin{eqnarray} 
\left \{ \left [ {\cal Y}(\eta,T,\mu) 
+ \eta \frac {d}{d \eta} {\cal Y}(\eta,T,\mu) \right ]   
\left[ \eta - \sigma_c +
\eta \frac{\lambda}{2 \pi N} {\cal Y}(\eta,T,\mu) \right]
+ \frac{\lambda T^2}{2 \pi N}J_2(\eta/T,\mu/T)
\frac {d}{d \eta}J_2(\eta/T,\mu/T)
\right\}\Bigr|_{\eta = \bar{\eta}} = 0 \;,
\label{genpms} 
\end{eqnarray} 
where we have defined the function
 
\begin{equation} 
{\cal Y}(\eta,T,\mu) =\ln \left ( \frac {M}{\eta} \right ) -  
I_2(\eta/T,\mu/T) \;. 
\end{equation} 
 
\noindent
Eq. (\ref {genpms}) expresses our general PMS result, Eq.
(\ref{pmsselfconsistent}), for the specific $d=2$ case. This can be easily
seem by recalling that in this number of space dimensions (and $\delta = 1$)
the exchange (Fock) type of self energy is given by
\begin{equation}
\Sigma_a(\eta,T,\mu) = -\eta \frac{\lambda}{2 \pi N} {\cal Y}(\eta,T,\mu)
\end{equation}

In the following we will present the analytical (whenever possible) and
numerical results for Eq. (\ref{genpms}). {}For convenience, the results will
be presented in units of $M$ for different values of $\lambda$ and $N$. We
start by analyzing the simplest case, of zero temperature and density.

\subsection{The $T=\mu=0$ case}

Taking Eq. (\ref {genpms}) at $T=\mu=0$ [that is, ${\cal
  Y}(\eta,T=0,\mu=0)=\ln (M/\eta)$] one gets
 
\begin{equation} 
\left [ \ln \left ( \frac {M}{{\bar \eta}} \right ) - 1 \right ]  
\left [ {\bar \eta} - \sigma_c - {\bar \eta} \frac {\lambda}{2 \pi N}  
\ln \left ( \frac {{\bar \eta}}{M} \right ) \right ]=0 \,\,. 
\label{pmstmuzero} 
\end{equation} 
 
\noindent 
As discussed previously, the first factor leads to the coupling independent
result, ${\bar \eta} = M/e$, which we shall neglect.  At the same time the
second factor in (\ref{pmstmuzero}) leads to a self-consistent gap equation
for $\bar \eta$, given by
 
\begin{equation} 
{\bar \eta}_{\delta^1} (\sigma_c)= \sigma_c \left[ 1- \frac {\lambda}{2 \pi N} 
 \ln \left ( \frac {{\bar \eta}_{\delta^1}}{M} \right )\right]^{-1} \;.
\label{etatzero} 
\end{equation} 
 
\noindent 
The solution for $\bar \eta_{\delta^1}$ obtained from Eq. (\ref{etatzero}) is

\begin{equation} 
\bar{\eta}_{\delta^1}(\sigma) =  -\frac{2 \pi N}{\lambda}\: 
W^{-1}\left[-\frac{2\pi N}{\lambda} \frac{\sigma_c}{M}\, \exp\left( 
-\frac{2 \pi N}{\lambda} \right)\right] \sigma \;, 
\label{etabarsol} 
\end{equation} 
where $W(x)$ is the Lambert $W$ (implicit) function \cite {lambert}, which
satisfies $W(x) \exp[W(x)] = x$.
 
To analyze CS breaking we then replace $\eta$ by Eq. (\ref{etabarsol}) in Eq.
(\ref {Vdelta1}), which is taken at $T=0$ and $\mu=0$.  As usual, CS breaking
appears when the effective potential displays minima at some particular value
${\bar \sigma_c} \ne 0$ which is obtained by minimizing the resulting
effective potential with respect to $\sigma_c$,

\begin{equation} 
 \frac{\partial V_{{\rm eff},\delta^1}(\sigma_c,\eta=\bar{\eta}_{\delta^1})}
{\partial \sigma_c}
\Bigr|_{\delta=1,\sigma_c=\bar{\sigma}_c} = \frac{\bar{\sigma}_c}{\lambda}  
+\frac{1}{\pi} \;{\bar \eta} \ln \frac{{\bar \eta}}{M}\;
= 0\;. 
\label{dVeffsig} 
\end{equation} 
Since $m_F=\bar{\sigma}_c$, after some algebraic manipulation of Eq.
(\ref{dVeffsig}) using the definition of the $W(x)$ function, we find
 
\begin{equation} 
m_{F,\delta^1} (0) = \bar{\sigma}_c (T=0,\mu=0)=
M {\cal F} (\lambda,N)\left( 1 - \frac {1}{2N} \right)^{-1}\;, 
\label{allmf} 
\end{equation} 
where we have defined the quantity ${\cal F}(\lambda,N)$ as
 
\begin{equation} 
{\cal F}(\lambda,N)= \exp \left \{ -\frac {\pi}{\lambda[1-1/(2N)]}  
\right \}\;. 
\label{funcF} 
\end{equation} 

\begin{figure}[htb] 
  \vspace{0.5cm}
  \epsfig{figure=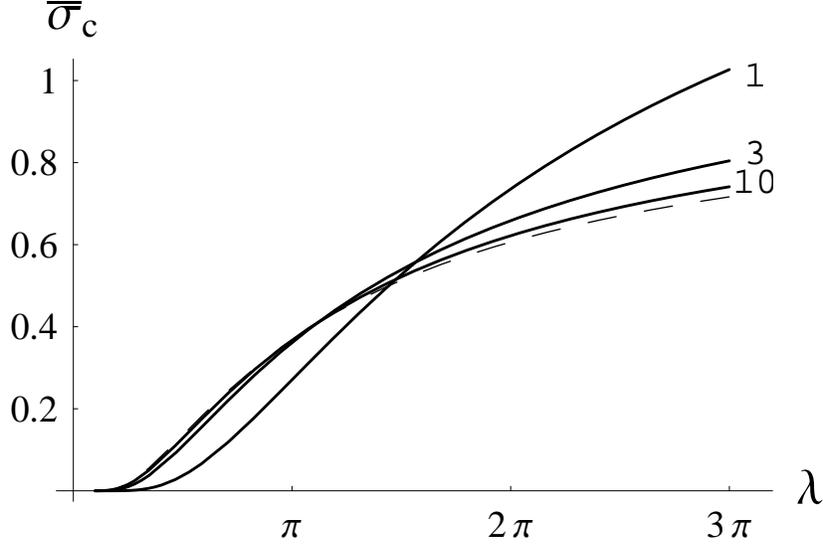,angle=0,width=12cm}
\caption[]{\label{alln} The dimensionless minimum ${\bar \sigma_c}$  
(in units of $M$)
  as a function of $\lambda$ for $T=\mu=0$. The dashed line represents the $N
  \to \infty$ result while the continuous lines were produced by the LDE-PMS
  at order-$\delta$. The numbers beside the curves identify the value of $N$
  for each case.}
\end{figure}

Eq. (\ref{allmf}) is our result for the fermionic mass at first order in
$\delta$ which goes beyond the large-$N$ result, Eq.(\ref{mF}).  Note also
that in the $N \to \infty$ limit, since ${\cal F}(\lambda,N \to \infty) = \exp
( - \pi/\lambda)$, Eq. (\ref{allmf}) correctly reproduces, within the LDE
non-perturbative resummation, the large-$N$ result. In {}Fig. \ref{alln} we
compare the order-$\delta$ LDE-PMS results for $\bar{\sigma}_c$ with the one
provided by the large-$N$ approximation.
 
At the same time, $\bar{\eta}$ evaluated at $ {\bar \sigma_c}(T=0,\mu=0)$, to
first order in $\delta$, also follows analytically from Eqs. (\ref{etabarsol})
and (\ref{allmf}),
 
\begin{equation} 
\bar{\eta}_{\delta^1}(\bar{\sigma}_c) = \left( 1 - \frac {1}{2N} \right) \:
\bar{\sigma}_c = M {\cal F}(\lambda,N)\;.
\label{alleta} 
\end{equation} 
The analytical results Eqs. (\ref{allmf}) and (\ref{alleta}) are exact
expressions (of course, at this first order of the LDE) following only from
the minimization condition Eq. (\ref{dVeffsig}) after some algebra.  
In particular, we
emphasize the simple scaling relation obtained between $\bar\eta$ and
$\bar\sigma_c$ in Eq. (\ref{alleta}), that only depends on $N$, leading to the
result (\ref{allmf}), and that will prove to be a very good approximation even
for the more general case, $T \ne 0$, as we shall see next.  Note that
this relation is the explicit form, for this case $T=\mu=0$, of the more
general relation Eq. (\ref{selfcons}) above, namely:
$\Sigma_a(\mu=0,\eta,T=0)=-\sigma_c/(2N)$.\\

Actually, there is an alternative simpler way of deriving the results
(\ref{allmf})--(\ref{alleta}) without using the solution for $\bar\eta$ in Eq.
(\ref{etabarsol}), that will moreover prove useful later on when we shall
consider the more complicate situations with $\mu \ne 0$, or typically also if
we would consider higher LDE order contributions.  That is, instead of using
Eq. (\ref{dVeffsig}) as giving ${\bar \sigma}_c(\bar\eta)$, we can use it to
substitute the rather complicated logarithmic dependence $\ln\bar\eta/M$ in
Eq. (\ref{etatzero}), canceling out at the same time the $\lambda$ dependence,
and obtaining thus a very simple linear equation for ${\bar \sigma}_c
/\bar\eta$ which only depends on $N$, 

\begin{equation}
\frac{\bar\eta}{{\bar \sigma_c}} = \left( 1 + \frac{1}{2N}
\frac{{\bar \sigma}_c}{\bar\eta} \right)^{-1}\;,
\label{direct}
\end{equation} 
which immediately leads to the results (\ref{allmf})--(\ref{alleta}).  The
solution in Eq. (\ref{etabarsol}) is useful however, as it gives $\bar\eta$
for any $\sigma_c$ values, away from the minimum in Eq. (\ref{dVeffsig}).

\subsection{The $T \ne 0$, $\mu=0$ Case}

{}For the finite temperature but zero chemical potential case, still using Eq.
(\ref{genpms}), the optimized $\bar{\eta}$ is now determined by the solution
of the expression
 
\begin{eqnarray} 
\left \{ \left [ {\cal Y}(\eta,T,\mu=0) 
+ \eta \frac {d}{d \eta} {\cal Y}(\eta,T,\mu=0) \right ]   
\left[ \eta - \sigma_c +  
\eta \frac{\lambda}{2 \pi N} {\cal Y}(\eta,T,\mu=0) \right]  
\right\}\Bigr|_{\eta = \bar{\eta}} = 0 \;. 
\label{genpmsmu0} 
\end{eqnarray}

\noindent 
The solution coming from the first term in Eq. (\ref{genpmsmu0}), corresponds
to the unphysical optimized result, whereas the solution obtained from the
second term gives the equivalent of the self-consistent gap equation,
(\ref{etatzero}),
 
\begin{equation} 
{\bar \eta}_{\delta^1}(\sigma_c,T) = \sigma_c 
\left[ 1+ \frac {\lambda}{2\pi N} 
{\cal Y}(\bar{\eta},T,\mu=0) \right]^{-1}\;. 
\label{etatmu0} 
\end{equation}

\begin{figure}[htb] 
  \vspace{0.5cm}
  \epsfig{figure=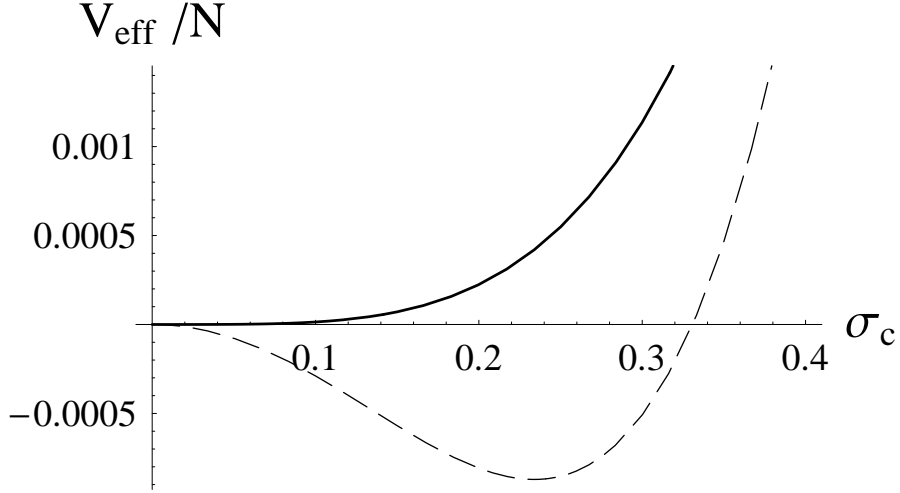,angle=0,width=12cm}
\caption[]{\label{veffhighT} The effective potential, $V_{\rm eff}/N$ as a
  function of $\sigma_c$ for $\mu=0$ and $T = 0.170757\,M$.  The parameter
  values are $\lambda=\pi$ and $N=3$.  The continuous curve, which shows CS
  restoration, was plotted using the LDE optimized results.  The dashed curve,
  which signals CS breaking, corresponds to the large-$N$ predictions. Both
  $V_{\rm eff}/N$ and $\sigma_c$ are in units of $M$.}
\end{figure}

\noindent 
We could next proceed by numerically solving Eq. (\ref{etatmu0}) for
$\bar{\eta}$ and substituting it in place of $\eta$ in Eq. (\ref{Vdelta1})
(evaluated at $\mu=0$). Then, by minimizing the effective potential we would
obtain the general behavior of ${\bar \sigma}_c$ as a function of the
temperature.  The critical temperature for CS restoration would then be
determined by the solution of ${\bar \sigma}_c(T=T_c)=0$ as usual.  However,
here an explicit analytical result for $T_c$ can also be obtained if we apply
the high temperature approximation to Eq. (\ref{Vdelta1}), with $\eta/T\ll 1$,
and then optimize the resulting expression.  The validity of using the
high-$T$ approximation before the optimization procedure could be questioned
in principle since $\eta$, at the level of Eq. (\ref{Vdelta1}), is arbitrary.
However, one may easily perform a cross check by performing a numerical PMS
application, as described above, without using the high-$T$ expansion. The
results we found for $T_c$ from both approaches agree very well with each
other, showing that the high-$T$ expansion is valid in a large range, though
actually $\bar \eta$ and $T_c$ are numerically of the same order of magnitude.
A simple reason for this is that the true high temperature expansion parameter
is not $\eta^2/T^2$, but rather $7/4 [\zeta(3)/(2\pi)^2] \eta^2/T^2 \sim 0.05
\eta^2/T^2$, as it is clear from Eq. (\ref{Vdelta1hit}) below, so that even if
$T_c \sim \eta$ the expansion parameter is still small enough.  If we then
expand Eq. (\ref{Vdelta1}) at high temperatures, up to order $\eta^2/T^2$, we
obtain
 
\begin{eqnarray} 
\frac{V_{{\rm eff},\delta^1}}{N} (\sigma_c, \eta, T, \mu=0) &=&
\delta \frac {\sigma_c^2}{2 \lambda} - T^2 \frac {\pi}{6} - 
 \frac{\eta^2}{2\pi}  \left [ \ln \left ( 
\frac {M e^{\gamma_E}}{T \pi} \right )  - \frac{ 7 \zeta(3)}{4 (2\pi)^2} 
\frac{\eta^2}{T^2} \right ] 
\nonumber \\ 
&+& \delta \frac{\eta(\eta-\sigma_c)}{\pi} \left [ \ln \left ( 
\frac {M e^{\gamma_E}}{T \pi} \right )  - \frac{ 7 \zeta(3)}{2 (2\pi)^2}  
\frac {\eta^2} 
{T^2} \right ] 
\nonumber \\ 
&+& \delta \frac {\lambda}{N} \frac {\eta^2 }{(2\pi)^2} 
 \left [ \ln^2 \left ( 
\frac {M e^{\gamma_E}}{T \pi} \right )  - \frac{ 7 \zeta(3)}{ (2\pi)^2}  
\ln \left ( 
\frac {M e^{\gamma_E}}{T \pi} \right ) \frac{\eta^2}{T^2} +  
{\cal O}(\eta^4/T^4)   \right ] \,\,.
\label{Vdelta1hit} 
\end{eqnarray} 
Then, setting $\delta=1$ and applying the PMS one gets
 
\begin{equation} 
\left [ \ln \left ( 
\frac {M e^{\gamma_E}}{T \pi} \right )  - \frac{ 21 \zeta(3)}{2 (2\pi)^2}  
\frac {{\bar \eta}^2} 
{T^2} \right ]\left \{ {\bar \eta} - \sigma_c + {\bar \eta}  
\frac {\lambda}{N(2\pi)} \left [ \ln \left ( 
\frac {M e^{\gamma_E}}{T \pi} \right )  - \frac{ 7 \zeta(3)}{2 (2\pi)^2}  
\frac {{\bar \eta}^2} 
{T^2} \right ] + {\cal O}({\bar \eta}^4/T^4) \right \}= 0  \,\,,
\label{pmsTmu0} 
\end{equation} 
where, once more, the first term in Eq. (\ref{pmsTmu0}) corresponds to the
unphysical ($\lambda$ independent) solution, while the second term gives
 
\begin{equation} 
{\bar \eta}(\sigma_c,T) =  
\sigma_c \left \{ 1+ \frac {\lambda}{N(2\pi)} \left [ \ln \left ( 
\frac {M e^{\gamma_E}}{T \pi} \right )  - \frac{ 7 \zeta(3)}{2 (2\pi)^2} \frac 
{{\bar \eta^2}} 
{T^2} \right ] \right \}^{-1} \,\,, 
\end{equation} 
whose solution can be expressed in the form

\begin{equation} 
{\bar \eta}(\sigma_c,T) = \sigma_c \left \{ 1+ \frac {\lambda}{N(2\pi)}  
\left [ \ln \left ( 
\frac {M e^{\gamma_E}}{T \pi} \right )  - \frac{ 7 \zeta(3)}{2 (2\pi)^2} \frac 
{\sigma_c^2} 
{T^2} \left [ 1+ \frac {\lambda}{N(2\pi)}  \ln \left ( 
\frac {M e^{\gamma_E}}{T \pi} \right ) \right ]^{-2}  +  
{\cal O}(\sigma_c^4/T^4) \right ] \right \}^{-1} \,\,. 
\label {etafinitet} 
\end{equation} 
The result given by Eq. (\ref{etafinitet}) is then plugged back into Eq. (\ref
{Vdelta1hit}) for $\delta=1$ and the resulting expression is again expanded
for $\sigma_c/T \ll 1$ in the high-$T$ approximation.  The order of the
transition can easily be checked numerically simply by plotting the effective
potential for different values of $T$ as shown in {}Fig. \ref{veffhighT}.

\noindent 
The result shows that the first order LDE result for finite $N$ predicts a
continuous phase transition as in the large-$N$ case.  By extremizing the
effective potential, at high-$T$, we obtain a maximum at ${\bar \sigma_c}=0$
and two minima at:
 
\begin{equation} 
{\bar \sigma_c}(T) = \pm \frac {T}{N^2 \sqrt{14 \pi \zeta(3)\lambda}} 
\left [ 2N \pi + \ln \left ( 
\frac {M e^{\gamma_E}}{T \pi} \right ) \right ]^{3/2}  
\left [ - 2N\pi+(2N-1) \lambda \ln \left ( 
\frac {M e^{\gamma_E}}{T \pi} \right ) \right ]^{1/2}\;. 
\label{sigmalde} 
\end{equation} 
 
\begin{figure}[htb] 
  \vspace{0.5cm}
  \epsfig{figure=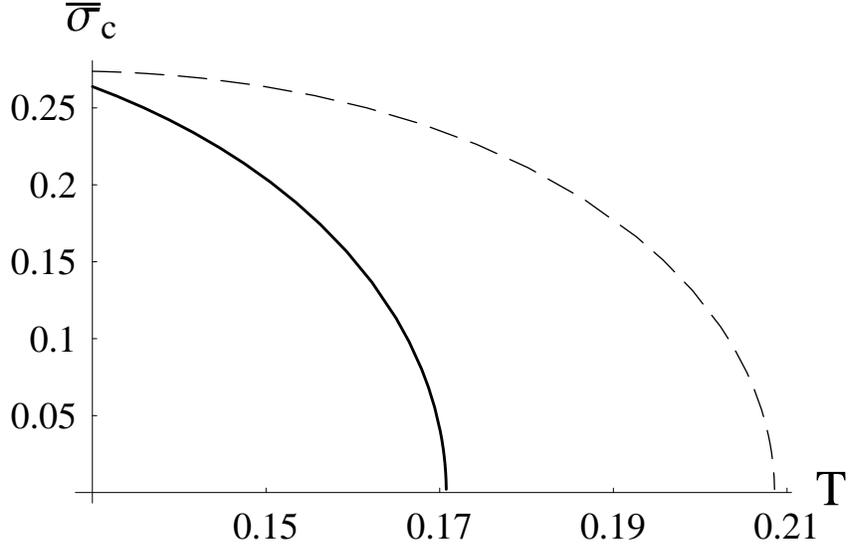,angle=0,width=12cm}
\caption[]{\label{mFhighT} The non-trivial  minimum
  ${\bar \sigma_c}$ as a function of $T$ for the parameter values
  $\lambda=\pi$ and $N=3$.  The first order LDE curve (continuous line)
  displays a continuous phase transition occurring at the critical temperature
  $T_c = 0.170 \, M$ while the large-$N$ result is $T_c = 0.208 \, M$. Both
  ${\bar \sigma}_c$ and $T$ are in units of $M$.}
\end{figure}

\noindent 
{}Figure \ref{mFhighT} shows ${\bar \sigma_c}/M$ given by Eq.
(\ref{sigmalde}) as a function of $T/M$, again showing a continuous (second
order) phase transition for CS breaking/restoration.

\begin{figure}[htb] 
  \vspace{0.5cm}
  \epsfig{figure=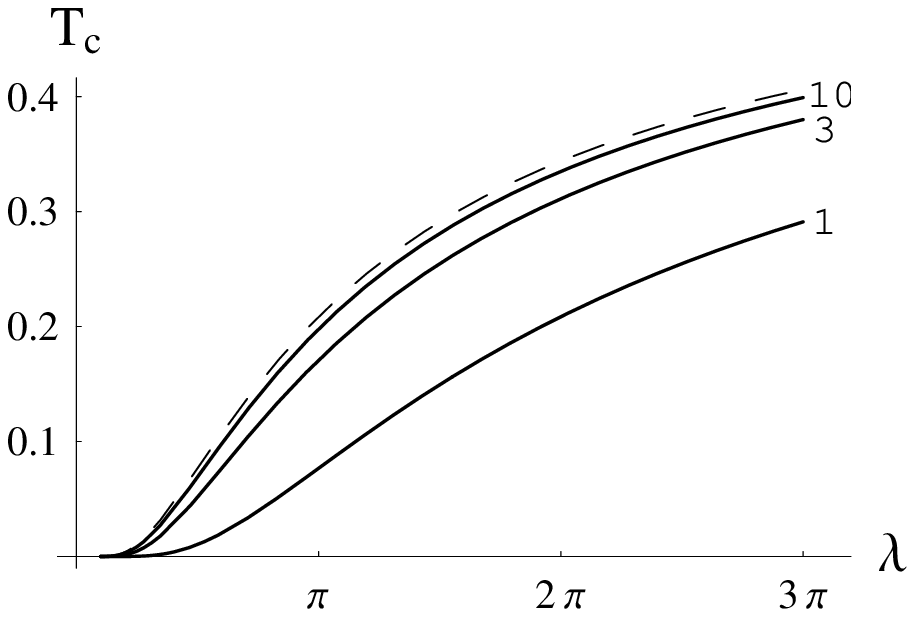,angle=0,width=12cm}
\caption[]{\label{allTc} The dimensionless critical temperature,
  $T_c$ (in units of $M$), as a function of $\lambda$ for $T\ne 0$ and
  $\mu=0$. The dashed line represents the $N \to \infty$ result while the
  continuous lines were produced by the LDE-PMS at order-$\delta$. The numbers
  next to the curves identify the value of $N$ for each case.}
\end{figure}

\noindent 
The numerical results illustrated by {}Figs. \ref{veffhighT} and \ref{mFhighT}
show that the transition is of the second kind and an analytical equation for
the critical temperature can be obtained by requiring that the minima vanish
at $T_c$. {}From Eq. (\ref {sigmalde}) one sees that ${\bar
  \sigma_c}(T=T_c)=0$ can lead to two possible solutions for $T_c$.  The one
coming from
 
\begin{equation} 
\left [ 2N \pi + \ln \left ( 
\frac {M e^{\gamma_E}}{T_c \pi} \right ) \right ] = 0\;, 
\end{equation} 
 
\noindent 
can easily be seen as not been able to reproduce the known large-$N$ result,
when $N\to \infty$, $T_c = M \exp(\gamma_E-\pi/\lambda)/\pi$.  However, the
other possible solution coming from
 
\begin{equation} 
\left [ - 2N\pi+(2N-1) \lambda \ln \left ( 
\frac {M e^{\gamma_E}}{\pi T_c } \right ) \right ] =0\;, 
\end{equation} 
gives for the critical temperature, evaluated at first order in $\delta$, the
result

\begin{equation} 
T_{c,\delta^1} = M \frac{e^{\gamma_E}}{\pi}   \exp 
\left \{ - \frac {\pi}{\lambda[1 -1/(2N)]} \right \} \, 
=\, M\frac{e^{\gamma_E}}{\pi} {\cal F}(\lambda,N) \;, 
\label{ldetc} 
\end{equation} 
with ${\cal F}_\lambda (N)$ as given before, by Eq. (\ref{funcF}).  Thus, Eq.
(\ref{ldetc}) exactly reproduces the large-$N$ result for $N \to \infty$.  The
results given by this equation are plotted in {}Fig.  \ref{allTc} as a
function of $\lambda$ for different values of $N$.

As already mentioned, the critical temperature predicted by Eq. (\ref{ldetc})
coincides, to a very good approximation, with the one observed in the
numerical results starting from Eqs. (\ref{Vdelta1}) and (\ref{genpmsmu0})
without the use of the high-$T$ approximation. We note in Eq. (\ref {ldetc})
the very same scaling relation than the one for $m_F(0)$ at zero temperature,
in Eq. (\ref{alleta}), thus involving a relation that only depends on $N$.
This simple scaling derives in fact from the equation for the optimized
solution $\bar\eta(T)$, where to obtain the factorized form Eq.
(\ref{pmsTmu0}), we have neglected ${\cal O}(1/T^4)$ small terms. Plugging
back these neglected terms, and manipulating Eq.  (\ref{pmsTmu0}), one easily
obtains the relation:

\begin{equation}
\sigma_c = {\bar \eta(T)} \left[ \left( 1-\frac{1}{2N}\right)^{-1} +
x^2 \frac{\lambda}{\pi} \,
H({\bar \eta},N)\right]\;,
\label{Tscaling}
\end{equation}
where we defined for convenience the high-temperature expansion parameter $x
\equiv 7/4 [\zeta(3)/(2\pi)^2] {\bar \eta}^2/T^2 \sim 0.05 \, {\bar
  \eta}^2/T^2$, so that $x^2 (\lambda/\pi) {\bar \eta} H({\bar \eta},N)$
defines the remnant part of order $1/T^4$, whose explicit form we do not need
to specify. Thus, neglecting these ${\cal O}(x^2)$ terms in Eq.
(\ref{Tscaling}) gives the very same relation between $\bar\eta$ and
$\sigma_c$ than at $T=0$ in Eq. (\ref{alleta}), and furthermore, Eq.
(\ref{Tscaling}) is directly related, after some algebra, to the result (\ref
{ldetc}) for $T_c$ above. Accordingly, though the simple scaling relation is
generally not expected to hold at arbitrary $T\ne 0$, the deviation from this
relation is essentially negligible due to the smallness of those ${\cal
  O}(x^2)$ corrections.  We shall argue later on, in section V.E, that this
universal scaling property is expected to remain a good approximation as well
at higher orders of the LDE approximation.

The (non-perturbative) LDE result shows that $T_c$ is always smaller (for the
realistic finite $N$ case) than the value predicted by the large-$N$
approximation. In the light of Landau's theorem for phase transitions in one
space dimensions, which predicts $T_c=0$, our LDE results, including the first
$1/N$ corrections, seem to converge to the right direction.

\subsection{ The $T=0$, $\mu \ne 0$ Case} 
 
As discussed in Ref. \cite {gnpolymers} this extremum of the phase diagram is
the very sensitive to the role played by kink like configurations. However, in
the present work only homogeneous background fields are considered and we are
not in position to compare our results for this part of the phase diagram with
the ones provided in that reference. Nevertheless, we are in position to
contribute by computing finite $N$ corrections so that one may, eventually,
use the LDE-PMS in conjunction with inhomogeneous background fields to further
improve the phase diagram found by the authors of Ref. \cite {gnpolymers}.
The case of zero temperature but finite chemical potential (density) also
follows from Eq. (\ref{genpms}). Using Eqs. (\ref{J2T0}) and (\ref{I2T0}) in
Eq. (\ref{genpms}), we find two situations. In the first, for $\eta > \mu$,
the optimized $\bar \eta$ is found from the solution of
 
\begin{equation} 
\left \{ \left [ \ln \left ( \frac{M}{\eta} \right ) - 1 \right ]  
\left [ {\eta} - \sigma_c + \frac {\lambda \eta}{2 \pi N}  
\ln \left ( \frac{M}{\eta} \right ) \right ]  \right \}
\Bigr|_{\eta=\bar \eta} = 0 \,.
\label{pmstmu} 
\end{equation}
{}For $\eta < \mu$, using the relations (\ref{J2T0}), (\ref{I2T0}) and
(\ref{JJ2T0}) in Eq. (\ref{Vdelta1}) we obtain that the optimized $\bar \eta$
is the solution of

\begin{equation}
\left\{
\left[ \eta-\sigma_c-\frac{\lambda \eta}{2 \pi N}
\ln  \left( {\frac {\mu+\sqrt {{\mu}^{2}-{\eta}^{2}}}{M}} \right) \right]
\left[ - \ln  \left( {\frac {\mu+\sqrt {{\mu}^{2}-{\eta}
^{2}}}{M}} \right) - \frac{\eta^2}{(\eta^2 -\mu^2 - \mu \sqrt {\mu^2- \eta^2})} \right]
- \frac{\lambda \eta }{2 \pi N}
\right\}
\Bigr|_{\eta=\bar \eta}  =0\;.
\label{mupms} 
\end{equation}
The solution for Eq. (\ref{pmstmu}) is exactly the one obtained previously,
given by Eq. (\ref{etabarsol}). Concerning Eq. (\ref{mupms}) one can again
obtain an analytical solution, by following a reasoning similar to the one
done for $T=\mu=0$ leading directly to Eq. (\ref{direct}), but with
slightly more involved algebra in the present case.  Thus, consider the 
non-trivial minimum ${\bar \sigma}_c$, obtained from 
$\partial V_{\rm eff}/\partial\sigma_c =0$,
now for $\mu \ne 0$. {}From Eq. (\ref{Vdelta1}) it follows that
${\bar \sigma}_c$ is given by 

\begin{equation}
{\bar \sigma}_c =
-\frac{\lambda}{\pi} \;\bar \eta \ln \left [\frac{\mu+\sqrt{\mu^2-\bar\eta^2}}{M} \right ]\;,
\end{equation}

\noindent
which replaces Eq. (\ref{dVeffsig}) for $\mu \ne 0$. Now again we can use this
to eliminate simply the complicated logarithmic dependence
$\ln[(\mu+\sqrt{\mu^2-\bar\eta^2})/M]$ in Eq. (\ref{mupms}), thus obtaining after
straightforward algebra a second order equation for ${\bar \sigma}_c$ as a
function of $\bar\eta$ and $\mu$, whose explicit solution reads:

\begin{equation}
{\bar \sigma}_c =  \frac{1}{2}\:\frac{\bar\eta}{1-1/(2N)}\:
\left[ 1+G(\lambda,\bar\eta,\mu,N) 
+\sqrt{[1-G(\lambda,\bar\eta,\mu,N)]^2-\frac{2}{N} 
\left(1-\frac{1}{2N}\right)\:
\frac{\lambda^2}{\pi^2}}
\;\;\right]\;,
\label{scalemu}
\end{equation}
where

\begin{equation}
G(\lambda,\bar\eta,\mu,N) \equiv 
\frac{\lambda}{\pi} \left(1-\frac{1}{2N} \right)\:
\left( 1-\frac{\mu}{ \sqrt{\mu^2-\bar\eta^2}} \right)\;,
\end{equation}

\noindent
contains the $\mu$ dependence. The relation (\ref{scalemu}) is the appropriate
generalization, for $\mu\ne0$, of the simple scaling relation obtained at
$T=\mu=0$ in Eq. (\ref{alleta}).  We have eliminated the other possible
solution for ${\bar \sigma}_c$ (namely with $\sqrt{\cdots} \to -\sqrt{\cdots}$
in Eq.  (\ref{scalemu})), by noting that for $\lambda\to 0$, Eq.
(\ref{scalemu}) correctly reproduces the simpler scaling relation in
(\ref{alleta}), while the other solution would give ${\bar \sigma}_c \to
0$.\footnote{The leading order large $N$ relation ${\bar \sigma}_c=\bar\eta$
  is also consistently reproduced for $N\to\infty$ in Eq. (\ref{scalemu}).}
It will prove also useful to expand (\ref{scalemu}) in powers of $\lambda$:

\begin{equation}
{\bar \sigma}_c \sim  \frac{\bar\eta}{1-1/(2N)}\:
\left[ 1 -\frac{1}{2N}\left (1- \frac{1}{2N} \right )\left (\frac{\lambda}{\pi} \right )^2
+\frac{1}{2N}\left (1- \frac{1}{2N} \right )^2\:
\left( 1-\frac{\mu}{ \sqrt{\mu^2-\bar\eta^2}} \right) \left (\frac{\lambda}{\pi} \right )^3+{\cal O}\left (\frac{\lambda}{\pi} \right )^4  \right]\;,
\label{expsiglambda}
\end{equation}

\noindent
which should be valid thus for moderate values of $\lambda/\pi$.  This
immediately shows that the corrections due to $\mu \ne 0$ to the simple
scaling obtained previously for $T=\mu=0$ in Eq. (\ref{alleta}) are actually
suppressed by ${\cal O}(\lambda^2/(\pi^2 N))$, moreover the $\mu$ dependence
enters only at the next order, $(\lambda/\pi)^3$.  These properties are
somewhat analogous to the case of the $T\ne0$ ($\mu=0$) corrections to the
scaling in Eq. (\ref{Tscaling}) which are also suppressed by the small
high-temperature expansion parameter.  We shall come back in the next
sub-sections to the important consequences of these relations for the general
case $T,\mu \ne 0$.

Next, we can extract the critical chemical potential, $\mu_c$, in an analogous
way as in the large-$N$ problem shown in Sub-sec. \ref{T0mufinite} and, as in
the large-$N$ problem, we can check that a first order transition also occurs.
To this aim we first calculate the effective potential at the value $\sigma_c$
given by Eq. (\ref{allmf}), where the relevant expression is given by $V_{\rm
  eff}(\sigma_c) $ for $\mu < \eta$. After some algebra many terms cancel out
so that we obtain simply

\begin{equation}
V^{\delta^{(1)}}_{\rm eff}(\sigma_c={\bar \sigma}_c,\eta=\bar\eta) =
-\frac{\bar\eta^2_{\delta^1}}{4\pi}\;,
\end{equation}

\noindent
{\it i.e.}, the same expression as the leading order $N\to\infty$ one, but
with the appropriate fermion mass at first order in $\delta$,
$\bar\eta_{\delta^1}$ given by Eq. (\ref{alleta}).  This has now to be
compared with the value of $V_{\rm eff}$ for $\mu\ne 0$ but with $\eta=0$,
which is simply obtained from Eq.  (\ref{Vdelta1}) as

\begin{equation}
V^{\delta^{(1)}}_{\rm eff}(\sigma_c=\eta=0,\mu) =
-\frac{\mu^2}{2\pi}\;\left(1-\frac{\lambda}{2\pi N}\right)\;,
\end{equation}
so that we can deduce an analytic expression for the critical density $\mu_c$,

\begin{equation}
\mu_{c,\delta^1} = \frac{\bar\eta_{\delta^1}}{\sqrt{2}} 
\left(1-\frac{\lambda}{2\pi N}\right)^{-1/2} \;,
\label{muc1}
\end{equation}

\noindent
valid at this first order in $\delta$.  The appearance of a pole in Eq.
(\ref{muc1}) for $\lambda = 2\pi N$ is an artifact of our first order in
$\delta$ approximation and thus probably not physically relevant.  
The point is that going to higher orders in the expansion in $\delta$
it will also bring different corrections of same order in powers of $1/N$, 
if expanded. Thus, we could for example limit ourselves in this analysis to 
the first $1/N$ order in this expansion, since complete
$1/N^2$ corrections are not included at the order in $\delta$ we are
considering. Seen this way, we could also expand Eq. (\ref{muc1}) 
in powers of $1/N$,

\begin{equation}
\mu_{c,\delta^1, 1/N} =  \frac{M}{\sqrt{2}} \; \exp(-\pi/\lambda)\:
\left[1 -\frac{\pi}{2 N\lambda} +\frac{\lambda}{4\pi N}
+{\cal O}(1/N^2) \right]\;,
\end{equation}

\noindent
which then exhibits no pole.

\begin{figure}[htb] 
  \vspace{0.5cm}
  \epsfig{figure=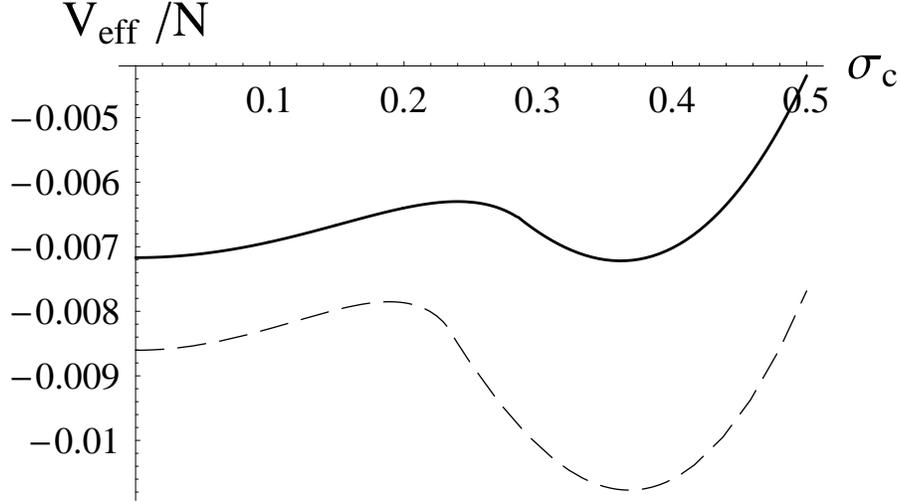,angle=0,width=12cm}
\caption[]{\label{veffmutzero}Large-$N$ (dashed line) and first order LDE  
  (continuous line) results for the effective potential, $V_{\rm eff}/N$. The
  parameter values are $N=3$, $\lambda=\pi$ and $T=0$. The effective potential
  has been evaluated at the (first order) LDE critical value $\mu_c=0.232 \,
  M$ for which the large-$N$ approximation still predicts CSB.  Both $V_{\rm
    eff}/N$ and $\mu$ are in units of $M$.  }
\end{figure}

We also proceeded numerically to obtain the solution from (\ref{mupms}) and
subsequent values of $\mu_c$, in order to have a useful crosscheck of the more
complicated most general case with both $T$ and $\mu$ finite, where analytical
expressions are not available.  An example for fixed values of $N$ and
$\lambda$ is presented in {}Fig.  \ref{veffmutzero}, which shows the effective
potential as a function of $\sigma_c$. The analogous condition to Eq.
(\ref{V=V}) leads to the result $\mu_{c,\delta^1} \simeq 0.232 \, M$, in the
first order of the LDE, that should be compared with the large-$N$ result,
$\mu_{c,{N\to\infty}} \simeq 0.260 \, M$.

\subsection{ The $T \ne 0$, $\mu \ne 0$ Case}

\begin{figure}[htb] 
  \vspace{0.5cm}
  \epsfig{figure=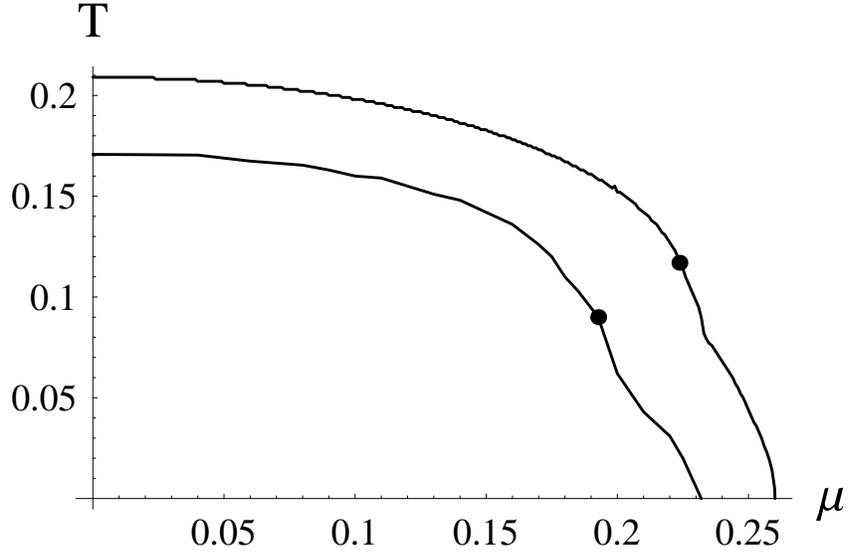,angle=0,width=12cm}
\caption[]{\label{phaselde} Large-$N$ (outmost line) and LDE (innermost line)  
  phase diagrams. The parameter values are $N=3$ and $\lambda=\pi$.  The dots
  represent the tricritical points.  The large-$N$ result for the tricritical
  point is $P_{tc,N\to \infty}=(T_{tc},\mu_{tc})=(0.117, 0.224)$ while
  the LDE approximation gives $P_{tc,\delta^1}\simeq
  (T_{tc},\mu_{tc})_{\delta^1} =(0.091, 0.192)$.  In both cases, the lines
  above the tricritical point represent second order phase transitions while
  the ones situated below represent first order transitions. All quantities
  are given in units of $M$.}
\end{figure}

We now arrive at the point of analyzing the complete LDE phase diagram, at
order-$\delta$. In the lack of possible fully analytical solutions in this
general case, we use numerical routines to determine both curves of second
order and first order phase transitions and their point of intersection which
then defines the tricritical point.  In {}Fig. \ref{phaselde} we show the
large-$N$ result compared to the LDE first order result. It can be seen that
in the LDE non-perturbative approach the region for CSB is diminished in an
appreciable way.  In units of $M$, the LDE result for the tricritical point,
with $N=3$ and $\lambda=\pi$, is $P_{tc,\delta^1}=(T_{tc},\mu_{tc})_{\delta^1}
=(0.091, 0.192)$ while the large-$N$ approximation gives $P_{tc,N\to
  \infty}=(T_{tc},\mu_{tc})=(0.117, 0.224)$.
 
\begin{figure}[htb] 
  \vspace{0.5cm}
  \epsfig{figure=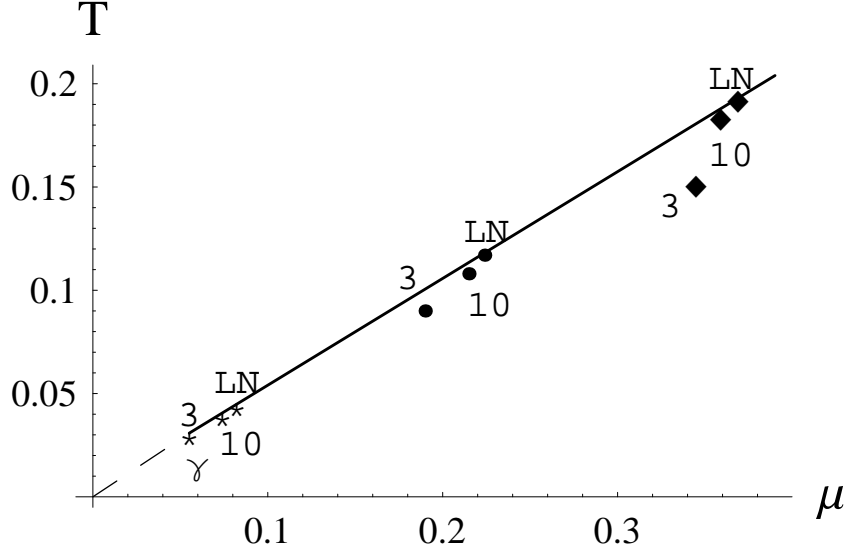,angle=0,width=12cm}
\caption[]{\label{alltricritical} The tricritical points 
  $P_t=(T_{tc},\mu_{tc})$, in units of $M$, for different values of $\lambda$
  and $N$.  The stars represent $\lambda=\pi/2$, the dots $\lambda=\pi$ and
  the diamonds $\lambda=2\pi$.  The corresponding values of $N$ are directly
  specified by the numbers on the graph.  The angle $\gamma$ represents the
  quantity $T_{tc}/\mu_{tc}$ as discussed in the text.  }
\end{figure}

It is worth remarking how the LDE tricritical point falls, approximately, over
a line joining the large-$N$ result and the origin. This remarkable result is
shown in {}Fig.  \ref{alltricritical} that shows $P_{tc}$ for $\lambda =
\pi/2, \pi,2\pi$ and $N=3,10$, and $N\to \infty$ (LN). A deviation of about $9\%$ is observed only for large values of the ratio $\lambda/N$ (e.g., $2\pi/3$). For ratios close to the unity the deviation is very small (about $3 \%$).

\subsection{Generality of Scaling for the LDE Results}

For moderate values of $\lambda/N$, we thus note the almost invariance, 
up to very small corrections in the LDE finite $N$
case, of the ratio $T_{tc}/\mu_{tc}$ that defines the angle $\gamma$,
 
\begin{equation} 
\tan \gamma = \frac {T_{tc}}{\mu_{tc}} \simeq 0.523 \;. 
\label{gamma} 
\end{equation} 
All our results show that this ratio remains to a good approximation
largely independent of both $N$ and $\lambda$. As already discussed,
this strongly suggests to postulate a very simple (but approximate) result for
the predictions of the tricritical points and for all other dimensionful
quantities obtained for the GN model within the LDE.  The result (\ref{gamma})
indicates that $T_{tc}$ and $\mu_{tc}$, as a function of $N$ and $\lambda$, to
a good approximation in the relevant range, scale according to
 
\begin{eqnarray} 
&& T_{tc} (\lambda, N) \simeq c\; g(\lambda, N)\, M \;,\\ 
&& \mu_{tc} (\lambda,N) \simeq g(\lambda, N) \,M \;,
\end{eqnarray} 
where $c= \tan \gamma$ and $g(\lambda, N)$ is a function of the parameters
$\lambda$ and $N$. At the first LDE order, we have obtained explicitly
$g(\lambda, N) = {\cal F}(\lambda,N)$, where ${\cal F}(N)$ is given by Eq.
(\ref{funcF}), up to very small corrections in the $T\ne0$ ($\mu=0$) case. For
the case $\mu\ne0$ ($T=0$) we saw that the approximation is valid only for
$\lambda \le \pi$ approximately, which is easily understood by examining the
approximate expanded form Eq. (\ref{expsiglambda}), of the relation Eq.
(\ref{scalemu}).  As anticipated, one can see that the corrections for $\mu\ne
0$ to the simple scaling relation (\ref{alleta}), strictly valid for $T=\mu=0$
only, are moderate as long as $\lambda$ is not too large, which remain
essentially true also for the more general case $T\ne0$ and $\mu\ne0$.

Interestingly, the actual result for the tricritical points, if compared with
the large-$N$ result shown in Subsec.  \ref{tricritsec}, is of the form
 
\begin{eqnarray} 
P_{tc}= (T_{tc},\mu_{tc})
\simeq (0.318, 0.608) \; \bar{\eta}_{\delta^1}(\bar{\sigma}_c)\;,
\label{tcfinal} 
\end{eqnarray} 
where $\bar{\eta}_{\delta^1}={\cal F}(\lambda,N) M$. The numerical deviations
from the above result are very small (less than $5\%$) for $\lambda/N < 1.3$.

The result (\ref{tcfinal}) together with our previous results obtained within
the first order in the LDE, Eqs. (\ref{allmf}), (\ref{ldetc}) and
(\ref{muc1}), respectively for the fermion mass (${\bar \sigma}_c$), critical
temperature ($T_{c}$) and critical chemical potential ($\mu_{c}$), show the
same approximate scaling as given in terms of the optimized LDE quantity
$\bar{\eta}_{\delta^1}$.  Also, as we have already seen previously, in the
large-$N$ limit, $\bar{\eta}_{\delta^1}(\bar{\sigma}_c) \to m_F(0)$ and all
our LDE results correctly reproduce those of Sec. \ref{basic}.

The reason why most of our results (except perhaps for the somewhat extreme
case $T=0,\mu\ne0$) exhibit approximately universal scaling properties as a
function of $\bar{\eta}$, to a very good approximation, are well understood at
the first LDE order, as explained in the previous subsections. On more general
grounds, in writing the interpolated Lagrangian density Eq. (\ref{GNlde}), a
explicitly dimensionful (mass) parameter, $\eta$, is introduced in the
originally massless model where there is no other dimensionful parameter.
Thus, from simple dimensional analysis, clearly all physical quantities should
scale with $\eta$, or actually $\bar{\eta}$ derived from the optimization
procedure.\footnote{Note that though at first the interpolation procedure done
  in Eq. (\ref{GNlde}) seems to explicitly break the discrete chiral symmetry
  of the model, $\eta$ is initially an undetermined parameter.  The
  introduction of the auxiliary field $\sigma$ just makes the optimized $\eta$
  a function of the background scalar field $\sigma_c$. After optimization
  $\bar{\eta}(\sigma_c) \sim \sigma_c$ and so, chiral symmetry breaking and
  restoration have the same interpretation as in the original
  (non-interpolated) GN model. It is a non-vanishing value for the minimum of
  the $\sigma$ field effective potential, $\langle \sigma_c \rangle =
  \bar{\sigma}_c \neq 0$ that still signals the breakdown of chiral symmetry,
  and $\bar{\eta}(\bar{\sigma}_c) \sim \bar{\sigma}_c$ provides the scale for
  all physical quantities derived.}  More precisely, at first LDE order, and
$T=\mu=0$, we obtained the exact simple relation Eq.  (\ref{alleta}) between
$\bar\eta$ and $\sigma_c$, which only depends on $N$.  More generally, in
terms of the basic scale of the model, say $\bar \Lambda \equiv M
e^{-\pi/\lambda}$ in the $\overline{\rm MS}$ renormalization scheme, truly
non-perturbative results are expected to give for the ratio of the relevant
quantities $m_F/\bar\Lambda$, $T_c/\bar\Lambda$ and $\mu_c/\bar\Lambda$, some
specific dimensionless coefficients depending only on $N$. A priori, there is
no reason why these coefficients should be exactly the same for the three
quantities. In contrast, on purely perturbative grounds, for $T\ne 0$ and $\mu
\ne 0$ we expect at finite LDE order, to obtain after optimization more
complicated scaling relations, {\it i.e.}, not only with different
coefficients for the three relevant quantities, but with such coefficients
being some non trivial (dimensionless) functions of $T$, $\mu$, $\sigma_c$,
$\lambda$, and $N$. Indeed, an explicit example that this is the case already
at first LDE order is illustrated by the relations Eq. (\ref{Tscaling}) for
$T\ne 0$, and Eq. (\ref{scalemu}) for $\mu\ne0$.  Both relations strictly
depend on $\lambda$, and $T$ (or, respectively: $\mu$).  However, we have seen
that, quite remarkably, this extra dependence upon the coupling is quite
suppressed, in such a way that the fermion mass $m_F$, critical temperature
$T_c$, and even $\mu_c$ have to a very good approximation identical scaling
factors which only depend on $N$.  Moreover, we have shown that in the $T\ne0$
($\mu=0$) case, this result is not a numerical accident but can be well
understood analytically by noting that the corrections to this universal
scaling are intrinsically quite negligible.  Even in the most general case
when both $T\ne0$ and $\mu\ne0$, the results for the tricritical point happens
in a regime where the simple scaling relation appears to be still a good
approximation. The only exception is the somewhat extreme case when $T=0$,
$\mu\ne$: in this case, if $\lambda$ is sufficiently large, the appropriate
generalization given analytically by Eq.  (\ref{scalemu}), leads to
substantial deviations from the simple scaling relation for $T=\mu=0$ in Eq.
(\ref{alleta}).

We believe in fact that this trend generalizes as well to arbitrary higher
orders of the LDE expansion, as we shall argue heuristically next.  To examine
what is happening at higher LDE orders, let us first consider
the $T=\mu=0$ case. It is simpler to generalize the
reasoning used to obtain Eq. (\ref{direct}) at first $\delta$ order, {\it
  i.e.}  using the solution $\sigma_c$ of Eq. (\ref{dVeffsig}) to replace the
logarithmic dependence $\ln ({\bar \eta}/{M}) $ into the (optimization)
equation defining $\bar\eta$, that thus generalizes Eq. (\ref{etatzero}) at
higher orders. In this way the latter will be a simpler, purely algebraic
equation, for the (only) parameter $\sigma_c/\bar\eta$, in terms of the
remaining parameters, the coupling $\lambda$ and $N$. More precisely, at
higher LDE orders, from general arguments the effective potential will have
additional perturbative terms of the generic form

\begin{equation}
\sim \sum_{k\ge 2} (\eta_*)^2 \delta^k \left (\frac{\lambda}{2\pi} \right )^k
\left[c_{LL}(N) \ln^{k+1}
\frac{\eta_*}{M}
+c_{NLL}(N) \ln^k\frac{\eta_*}{M}+\cdots \right]
\label{nextord}
\end {equation}  

\noindent
with leading, next-to-leading, etc different powers of $\ln(\eta_*/M)$, 
and $c_{LL}$, etc, the corresponding coefficients to be calculated from the 
relevant Feynman graphs. 
Next, performing the LDE to some given order and taking
the limit $\delta\to 1$ etc, the resulting dependence upon 
$\sigma_c$ and $\eta$ is much more involved than at first order,
so that one obtains from the CS breaking condition a
non-linear relation for $\sigma_c$, generalizing Eq. (\ref{alleta}), 
and as well for the optimized $\bar\eta$ solution,  
Eq. (\ref{etatzero}). Nevertheless, it is clear that 
these two relations can be used e.g. to eliminate the
logarithmic dependence $\ln\bar\eta/M$, thus obtaining a polynomial equation
for $\sigma_c/\bar\eta$, which only depends on $N$ and  
$\lambda$. 
Actually, if we were able to re-sum this LDE series to all orders, we would 
certainly expect that the only dependence on the coupling $\lambda$ in 
all physical quantities would be entirely included in terms 
of the basic dimensionful
scale of the model e.g. in the $\overline{\rm MS}$ scheme:
$ \Lambda_{\overline{\rm MS}} \equiv M e^{-\pi/\lambda}$.
However, at finite LDE orders, the optimization generally defines 
a rather complicated dependence on $\lambda$ for $\bar\eta$.  
But we have checked explicitly at the next ($\delta^2$) order
that this induces relatively small deviations from the first order
scaling relation in Eq. (\ref{alleta}) (upon assuming obviously that 
the unknown coefficients $c_{i}$ appearing in Eq. (\ref{nextord}) take 
generic values of ${\cal O}(1)$ ). 
Next, we saw that  the $T\ne0$ result gives,
at this  LDE order, a very small correction to the scaling
relation existing for  
$T=0$. For $\mu\ne0$, a deviation to this simple scaling
is predicted by Eq. (\ref{scalemu}), but it remains, however,
numerically moderate up to relatively large values of the coupling $\lambda$.
Now, since as we just examined the $T=0$ scaling properties are expected 
to generalize at higher LDE  
orders, and the LDE  generally converges quite rapidly,
we can expect
that the higher order modifications for $T,\mu \ne 0$ 
to these first LDE order scaling properties should remain small corrections.

\section{Comparing the LDE and the $1/N$ results}

Let us compare, in this section, the LDE leading order results with the ones
given by the 1/$N$ approximation at leading order (LO) as well as next to the
leading order (NLO).  As already pointed out, the effective potential, at
$T=0$ and $\mu=0$, for the 2d Gross-Neveu model to the next to leading order 
(NLO)
was first evaluated by Root \cite{root}. The NLO correction to the fermionic
mass, at $T=\mu=0$, was explicitly evaluated by Forg\'{a}cs, Niedermayer and
Weisz \cite{forgacs}. Using a combination of the thermodynamic Bethe ansatz
and the $1/N$ effective potential at $T=\mu=0$, Chodos and Minakata
\cite{muc1n} were able to obtain the NLO correction for $\mu_c$.  Recently,
the authors in Ref. \cite{blaizot} computed the complete NLO
in the $1/N$ expansion for the effective potential at $T \ne 0$ and $\mu = 0$
performing a detailed  numerical analysis of their results.  
They also exhibit in
details a number of non-trivial properties, in particular for the expected
behavior at high temperature, but because of the appearance of a Landau pole
near the $T \sim T_c$ regime, they do not conclude on a well-defined value of
$T_c$ from the full $1/N$ calculation. 
 It is thus difficult to compare our numerical LDE estimates of $T_c$
with their numerical results. 

Next, we emphasize that, as far as we know, there are no
$1/N$ NLO results for the case $T \ne 0$ and $\mu \ne 0$, thus no results for 
the
tricritical points beyond the large-$N$ approximation. {}For comparison
purposes, let us use our notations and conventions to present the only two
available analytical results.

{}For the case $T=\mu=0$ Ref. \cite{forgacs} gives\footnote{In our comparison 
with the exact $1/N$ result of the fermion mass as given in Ref.\cite{forgacs},
we should be cautious to {\em remove} from their expression a term: 
$M \exp(-\pi/\lambda)/N$, due our scheme being such that our 
 reference scale is $M \exp(-\pi/\lambda)$, rather than the full 
 $\Lambda_{\overline{\rm MS}}$ expression of \cite{forgacs}.}
 
\begin{equation}
{\bar \sigma}_c^{1/N,NLO}(0)= M \exp(-\pi/\lambda)\left [ 1 +
\frac {1}{N}(2 \ln 2 -\gamma_E) \right ] \;.
\label{zeromassnlo}
\end{equation}

\begin{figure}[htb] 
  \vspace{0.5cm}
  \epsfig{figure=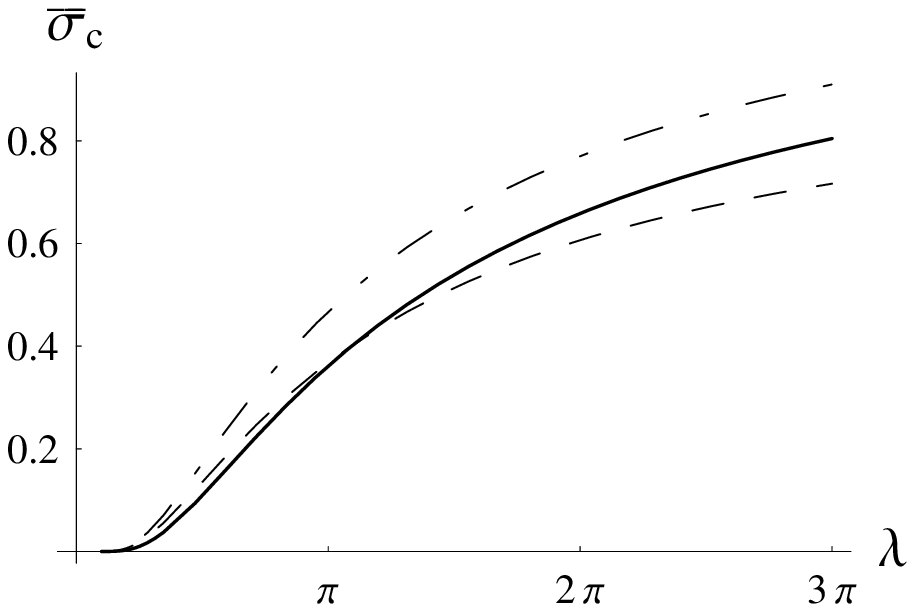,angle=0,width=12cm}
\caption[]{\label{gnfig12} The fermionic mass  $m_F(0)={\bar \sigma_c}$, in 
  units of $M$, plotted as a function of $\lambda$ for $N=3$, $T=0$ and
  $\mu=0$. The dashed line represents the $1/N$ result at leading order, the
  dot-dashed line represents the $1/N$ result at next to leading order and the
  continuous line is the first order LDE result.  }
\end{figure}

{}Figure \ref {gnfig12} shows the LDE fermionic mass at $T=0=\mu$, ${\bar
  \sigma}_c= M \,{\cal F}(\lambda,N)/[1-1/(2N)]$ as a function of $\lambda$ for
$N=3$. The same figure shows the $1/N$ results at LO, given by Eq. (\ref{mF}
), and at NLO, given by Eq. (\ref{zeromassnlo}).

{}For the case $T=0, \mu \ne 0$ Ref. \cite{muc1n} gives

\begin{equation}
\mu_c^{1/N, NLO} =\frac{M}{\sqrt 2} \exp(-\pi/\lambda)
\left( 1- \frac{0.47}{N} \right) \;.
\label{mucnlo}
\end{equation}
 
\begin{figure}[htb] 
  \vspace{0.5cm}
  \epsfig{figure=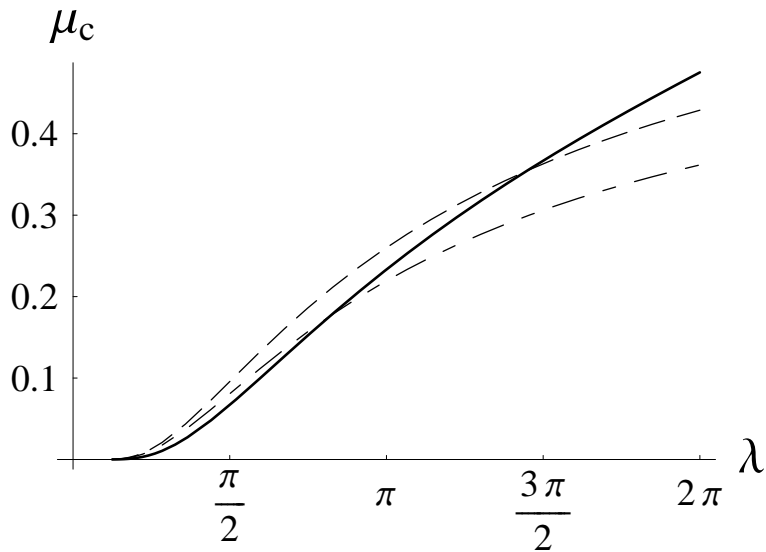,angle=0,width=12cm}
\caption[]{\label{gnfig13} The critical chemical potential $\mu_c$ in units 
  of $M$, plotted as a function of $\lambda$ for $N=3$ and $T=0$. The dashed
  line represents the $1/N$ result at leading order, the dot-dashed line
  represents the $1/N$ result at next to leading order and the continuous line
  is the first order LDE result.}
\end{figure}
{}Figure \ref{gnfig13} shows the LDE critical chemical potential at $T=0$,
${\mu}_c= M/\sqrt{2} \, {\cal F}(\lambda,N)$, as a function of
$\lambda$ for $N=3$. The same figure shows the $1/N$ results at LO, given by
Eq. (\ref{largeNmuc} ), and at NLO, given by Eq. (\ref{mucnlo}). As we have emphasized in the text, the LDE results for this case are valid only up to $\lambda \sim \pi$. Apart from that, this is the case where kink like configurations start to play a major role so any quantitative results must be interpreted with the due care \cite {gnpolymers}.

\section{Conclusions}

The analytical non-perturbative technique known as the LDE has been applied to
the two dimensional Gross-Neveu model effective potential at finite
temperature and chemical potential. {}Following the prescription suggested in
Ref. \cite{npb} we have shown that, within the large-$N$ limit, the LDE
exactly agrees with $1/N$ LO ``exact" result for any values of $T$ or $\mu$.
Having established this reliability we have considered the first finite $N$
correction that already appears at the first non-trivial order. The
variational optimization procedure has produced interesting results that
turned out to be possible to be cast into analytical form.  A careful
analysis of our numerical and analytical optimized results has led us to write
down five simple relations, that take into account finite $N$ corrections,
concerning the $T=0=\mu$ fermionic mass (${\bar \sigma}_c$), $\mu_c$ (at
$T=0$), $T_c$ (at $\mu=0$), and the tricritical points, ($T_{t,c},\mu_{t,c}$).
As we have discussed, four of these quantities essentially scale, 
to a very good
approximation, with the LDE optimized mass scale at $T=0=\mu$ given by
$\bar{\eta}_{\delta^1}/M={\cal F}(\lambda,N)$, with ${\cal F}(\lambda,N)$
defined by Eq. (\ref{funcF}), with $\bar{\eta}$ being the only dimensionful
quantity present in the interpolated Lagrangian density. 
The only exception is the case of $\mu_c$ for $T=0$ where, for large enough
coupling $\lambda$, we obtained substantial deviation from the simple scaling,
with an explicit expression of its dependence on $\lambda$. 
However, as already mentioned, we believe our results are not
faithful in the small $T$ and large $\mu$ part of the phase diagram, which
is the most substantially affected by inhomogeneous backgrounds 
as shown at leading $1/N$ order in Ref. \cite{gnpolymers}. The basic reasons for
the approximately very good validity of the universal scaling relations, at
the first LDE order, were identified and understood, as explained in some
detail in Sec. V. Moreover, we argued that this quasi-universality of scaling 
is expected to hold also at arbitrary higher orders of the LDE.  
At $T \ne 0$ and
$\mu \ne 0$ our main results concern the evaluation of the phase diagram,
containing finite $N$ corrections that, as far as we know, has not been
carried out before.  Comparing our perturbative type of evaluation with the
ones performed in Refs. \cite{forgacs,blaizot,muc1n}, for example, one may
notice some of the LDE advantages. Namely, it automatically introduces an
infrared cutoff that makes possible completely perturbative evaluations.  At
the same time, at each (perturbative) order one has just a few {}Feynman
graphs to evaluate as compared to the traditional non-perturbative methods,
such as the $1/N$ approximation. In particular, this advantage of the LDE
procedure means that the renormalization program can be easily implemented.
Now to proceed beyond the first LDE order, as far as the $T\neq 0$,
$\mu=0$ case is concerned, it could be interesting 
in principle to exploit those full $1/N$ results of Ref.
\cite{blaizot} , by re-expanding these in ordinary perturbation theory 
in the coupling $\lambda$, and then proceeding with the LDE procedure
as outlined in section III. However, after definite efforts to do this, 
it proves to be of very little use for our purposes, simply because beyond
the first LDE order one can obtain these results only
in purely numerical form. While, to perform the LDE, we necessarily need 
at least to have the analytical dependence upon the coupling $\lambda$
and fermion mass to be able to perform the interpolation as described
in Eq. (\ref{GNlde}). Moreover, even
at first order, it is very difficult to compare our results with theirs,
due to the different momenta routing used, resulting in their much
more involved expressions of the $T$-dependent contributions in particular
(note however that this alternative routing is the only possible one beyond
the first order).   
More generally, at the first LDE order here investigated, we cannot 
expect to get very close to the exact $1/N$
results,
but at the same time we expect the LDE to converge faster and it will
incorporates terms beyond the $1/N$ at higher orders, as discussed before.

Speculating on the expected behavior at higher LDE orders, 
the occurrence of a Landau pole 
in the complete $1/N$ results found by the authors of Ref. \cite{blaizot}
therefore invalidating a priori a unambiguous determination of a $T_c$ value, 
deserves some general comments. 
Accordingly, it is useful to recall how  the Landau pole
emerges in the construction of  Ref. \cite{blaizot}:
in very rough terms, at finite $T$ a pole can occur in the dressed 
propagator of the $\sigma$ field for some value of the 
$T$-dependent effective fermion mass, $m_F(T)$, because the latter decreases
as $T$ increases from zero, reaching eventually $m_F(T_c)=0$ at critical
temperature\footnote{Note in contrast that this Landau pole is harmless at 
$T=0$, because in the corresponding exact $1/N$ calculation of the mass 
gap (see e.g. \cite{forgacs,jlreno}), this pole actually
cannot be reached for any consistent value of the mass gap.}. 

Now, it is interesting to note that the LDE will avoid this Landau
pole problem, in a rather trivial way:  
since by construction the LDE stops at {\em finite} orders, 
where there cannot be a pole, while a Landau
pole is relevant only when considering a resummed 
perturbative series. Moreover, let us assume that one would manage
to re-sum the LDE perturbative
series to all orders, in some approximation (which can be done 
explicitly e.g. for the GN mass gap \cite{ldegn,jldamien} at order-$1/N$, using the LDE together with renormalization group properties):
one would thus obtain a resummed expression
exhibiting possibly a pole at a critical mass value, but even in such a case,   
the optimization prescription used in the LDE construction will escape 
this pole, i.e the 
optimization ``freezes out" the mass at a value which generally cannot 
coincide with a possible Landau pole value. 

Although convergence properties cannot be accessed in the first non-trivial
order, the fact that our $T_c$ is, in accordance with Landau's theorem,
smaller than the LO large $N$  result gives further support to the method.
However, a deeper discussion about convergence is beyond the scope of the
present work due to the technical difficulties in evaluating, at $T \ne 0$ and
$\mu \ne 0$, some of the three loop graphs shown in Fig. \ref {gnfig2}. We
recall that the LDE-PMS convergence properties in critical theories have been
analyzed, by the present authors, in connection with homogeneous Bose-Einstein
condensates \cite{knp}.  We intend to extend the present work by considering
the order-$\delta^2$ three loop graphs that will bring the first $1/N^2$
corrections. This will help us to gauge convergence properties, the effects of
the $1/N^2$ terms, as well as the eventual generalization of the universal
scaling relations between the critical and tricritical quantities with the LDE
optimized mass scale.

\acknowledgments
 
M.B.P. and R.O.R. are partially supported by CNPq-Brazil.  M.B.P.  thanks the
Laboratoire de Physique Th\'{e}orique et Astroparticules (Universit\'{e} de
Montpellier II) for a CNRS guest grant and R.O.R.  also acknowledges partial
support from FAPERJ. This work is dedicated to the memory of Philippe Garcia,
who loved science.

\appendix   

\section{Summing Matsubara frequencies and related formulas}

The integrals encountered in Feynman's graphs are performed, as usual,
at finite temperature and density with the substitution rules
$p_0 \to i(\omega_n  - i\mu)$ where $\mu$ is the chemical potential and
$\omega_n=(2 n +1) \pi T$, $n=0, \pm 1, \pm 2, \ldots$, are the Matsubara
frequencies for fermions.
We sum over the Matsubara frequencies with 
standard contour integration techniques, and regularize the remaining
(Euclidean) momentum integrals with dimensional
regularization  and  carry the renormalization in the
$\overline{\rm MS}$
scheme. 
E.g., momentum integrals of
functions $f(p_0,|{\bf p}|)$ are replaced by (see Ref.  \cite{kapusta})

\begin{eqnarray}
\int  \frac {d^2 p}{(2 \pi)^2} f(p_0,|{\bf p}|) \to
iT\sum \hspace{-0.5cm} \int \; f[i(\omega_n  - i\mu),|{\bf p}|] =
iT \sum_n \int  \frac {d p}{(2 \pi)}
\; f[i(\omega_n  - i\mu),|{\bf p}|]\;,
\nonumber
\end{eqnarray}
{}For the divergent, zero temperature
contributions, dimensional regularization is carried out in 
dimensions
$d= 1-2\epsilon$ and in the $\overline{\rm MS}$
scheme, in which case the momentum integrals are written as

\[
\int \frac {dp}{(2 \pi)} \to \int_p = \left(\frac{e^{\gamma_E} M^2}{4 \pi}
\right)^{\epsilon} \int \frac {d^{d} p}{(2 \pi)^{d}} \;,
\]

\noindent
where $M$ is an arbitrary mass scale and $\gamma_E \simeq 0.5772$ is the
Euler-Mascheroni constant.

The Matsubara sums which are relevant for the different integrals considered in
section IV can be derived as (see e.g. \cite{kapusta}):

\begin{equation}
T \sum_{n=-\infty}^{+\infty}
\ln [(\omega_n-i \mu)^2 + \omega^2({\bf p})]
= \omega({\bf p}) + T \ln\{1+ e^{-[\omega({\bf p})+\mu]/T}\} 
+ T \ln\{1+ e^{-[\omega({\bf p})-\mu]/T}\}\;,
\label{sum1}
\end{equation}

\begin{equation}
T \sum_{n=-\infty}^{+\infty}
\frac{1}{(\omega_n-i \mu)^2 + \omega^2({\bf p})}
= \frac{1}{2 \omega({\bf p})} \left\{ 1- \frac{1}{e^{[\omega({\bf p})+\mu]/T}+1}
- \frac{1}{e^{[\omega({\bf p})-\mu]/T}+1} \right\}\;,
\label{sum2}
\end{equation}

\begin{equation}
T \sum_{n=-\infty}^{+\infty}
\frac{\omega_n-i \mu}{(\omega_n-i \mu)^2 + \omega^2({\bf p})}
= \frac{i}{2}  \frac{\sinh (\mu/T) }{\cosh(\mu/T) +
\cosh( \omega({\bf p})/T ) }\;,
\label{sum3}
\end{equation}

where $\omega^2({\bf p})= {\bf p}^2 +\eta^2$.

\section{Renormalization of the LDE Effective Potential} 
 
We give here some details on the renormalization procedure for
the effective potential of the interpolated model (\ref{GNdelta}).
{}First, let us consider the non-renormalized result for the effective
potential in the large-$N$ approximation, which from Eq. (\ref{veffNlde}) is
given by
 
\begin{equation} 
\frac{V_{{\rm eff},\delta^1}}{N} (\sigma_c,\eta_*,N \to \infty)= 
\delta  \frac {\sigma_c^2}{2 \lambda} - 
\frac{1}{2\pi} \left \{ \eta_*^2 \left [ \frac {1}{\epsilon}+\frac {1}{2} + 
\ln \left( \frac {M}{\eta_*} \right ) \right] + 
2 T^2 I_1\left( \frac{\eta_*}{T},\frac{\mu}{T}\right) \right \} \;. 
\label{AVdeltadiv} 
\end{equation} 
 
\begin{figure}[htb] 
  \epsfig{figure=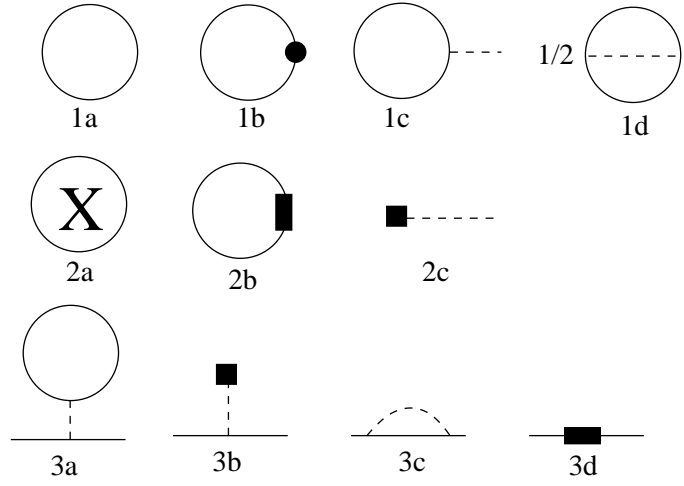,angle=0,width=9cm}
\caption[]{\label{figren} 
  Feynman diagrams contributing to the effective potential up to order
  $\delta$.  Diagrams 1a, 1b, 1c and 1d are the LDE ultra violet divergent
  contributions. Diagram 1d has a symmetry factor 1/2.  Diagram 2a represents
  the counterterm corresponding to the zero point energy subtraction. Diagram
  2b is constructed with the counterterm (3d) which renormalizes the exchange
  self energy (3c). The linear counterterm 2c can be obtained by renormalizing
  (with 3b) the direct self energy term, 3a. }
\end{figure}

Going beyond large-$N$ one must add the order-$\delta$ term, Eq. (\ref{VN1}),

\begin{equation} 
 V_{{\rm eff},\delta^1}^{(a)} (\eta_*)=
-i \frac {1}{2} \int \frac {d^d p}{(2\pi)^d} 
{\rm tr} \left [\frac {\Sigma_a(\eta_*)}{\not \! p - \eta_* + 
i \epsilon} \right ]\;, 
\label{AVN1} 
\end{equation} 
where $\Sigma_a$ is the term 3c in {}Fig. \ref{figren},

\begin{equation} 
\Sigma_a (\eta_*) = -\delta  \left (\frac {\lambda}{N} \right ) 
i\int \frac {d^d q}{(2 \pi)^d} \frac {1}{\not \! q - \eta_*+i \epsilon}\;. 
\label{ASigma1} 
\end{equation} 

It is worth noting that the relative
simplicity of our final result, Eq. (\ref{DVN1}), for this two-loop
graph is largely due to the appropriate
choice of momenta routing, allowing to factorize the
result into (squared) one-loop contributions. We checked explicitly
that it is strictly equivalent to another momenta routing choice
in the literature \cite{root,blaizot},
more appropriate when considering higher order contributions
but which  at this two-loop level would result into much more 
involved intermediate expressions.

 As it has been shown \cite{ldegn,prd1}, the standard $\overline{\rm MS}$ 
renormalization procedure and the LDE commute with each other, 
so that one may perform
the LDE before renormalization, introducing thus extra $\delta$-dependent
counterterms,
or alternatively directly on renormalized quantities. We shall rather 
follow here the first approach which may be more illustrative for our 
purpose. 
 An explicit evaluation of Eqs. (\ref{AVN1}) and (\ref{ASigma1}) 
with the rules given in appendix A gives
  
\begin{eqnarray} 
 \frac {V_{{\rm eff},\delta^1}^{(a)}}{N}(\sigma_c,\eta_*,T,\mu) &=&
\delta  \frac{\lambda}{4\pi^2 N} \left\{\eta^2_* 
\left \{ \frac{1}{4\:\epsilon^2} 
+ \frac{1}{\epsilon} \left [ \ln \left ( \frac {M}{\eta_*}  \right ) - 
I_2(\eta_*/T,\mu/T) \right ] \right . \right. \nonumber \\ 
&+& \left . \left. \left [ \ln \left ( \frac {M}{\eta_*}  \right ) - 
I_2(\eta_*/T,\mu/T) \right ]^2 +\ln^2 
 \frac {M}{\eta_*} + \frac{\pi^2}{24}\right \}
+T^2 J^2_2(\eta_*/T,\mu/T) \right\}
\;. 
\label{ADVN1div} 
\end{eqnarray} 
Eqs. (\ref{AVdeltadiv}) and (\ref{ADVN1div}) give the total contributions from
diagrams 1a,1b,1c and 1d in figure \ref {figren}.

To evaluate the contribution given by graph 2b in {}Fig. \ref{figren} we need
to define the mass counterterm $D^{\delta}_{exc}
{\bar \psi} \psi$ used to renormalize the exchange self energy given by 3c.
Note that the divergences are only associated to the terms proportional to the 
mass.  
This sets the Feynman rule to $iD^{\delta}_{exc}$ where

\begin{equation} 
D^{\delta}_{exc}= -\delta \frac{\lambda}{4 \pi N}  \eta 
\frac {1}{\epsilon} \;.
\end{equation} 
Then, one can evaluate the one loop contribution corresponding to 2b in {}Fig.
\ref{figren},

\begin{equation} 
 \frac{V_{{\rm eff},\delta^1}^{(a,cterm)}}{N} (\eta_*)=
-i  \int \frac {d^d p}{(2\pi)^d} 
{\rm tr} \left [\frac {D^{\delta}_{exc}}{\not \! p - \eta_* + 
i \epsilon} \right ]\;, 
\label{AVN1_2} 
\end{equation} 
obtaining

\begin{equation}
 \frac{V_{{\rm eff},\delta^1}^{(a,cterm)}}{N}(\sigma_c,\eta_*,T,\mu) =
 \delta  \frac{\lambda}{4\pi^2 N}\eta^2_* 
\left\{ -\frac{1}{2\:\epsilon^2} 
- \frac{1}{\epsilon} \left[ \ln \left ( \frac {M}{\eta_*}  \right ) - 
I_2(\eta_*/T,\mu/T) \right ]  
  -\ln^2 \frac {M}{\eta_*} - \frac{\pi^2}{24} \right\} 
\label{ADVN1div2} 
\end{equation}
so that when adding the two contributions (\ref{ADVN1div2}) and
(\ref{ADVN1div}) the $1/\epsilon$ divergence as well as some finite 
terms cancel
out, and there only remains the $1/\epsilon^2$ divergence to be renormalized
by zero point (vacuum energy) subtraction counterterms.

%
The {}Feynman renormalization coefficient corresponding to the linear
counterterm $E^{\delta}\sigma$ is obtained from the divergent part of the
fermionic loop contained in the graph 3a of {}Fig. \ref{figren}, namely

\begin{equation} 
E^{\delta} = \delta \frac{\eta}{\pi \epsilon}\;. 
\end{equation} 
Then, adding all contributions one has

\begin{eqnarray} 
\frac{V_{{\rm eff},\delta^1}}{N} (\sigma_c, \eta, T, \mu) &=&
\delta \frac {\sigma_c^2}{2 \lambda} - 
 \frac{1}{2\pi} \left \{ \eta^2 \left [ \frac {1}{2} + \ln \left ( 
\frac {M}{\eta} \right ) \right ] + 2 T^2 I_1(\eta/T,\mu/T) \right \} 
\nonumber \\ 
&+& \delta \frac{\eta(\eta-\sigma_c)}{\pi} 
\left[\ln\left(\frac{M}{\eta}\right) - I_2 (\eta/T,\mu/T) \right] 
\nonumber \\ 
&+& \delta \frac {\lambda}{N} \frac {\eta^2}{4\pi^2} 
  \left [ \ln \left ( \frac {M}{\eta}  \right ) - I_2(\eta/T,\mu/T) 
\right ]^2 \nonumber \\ 
&-& \frac{\eta^2}{2\pi \epsilon} + \frac{\delta}{\pi \epsilon} 
\left[\eta^2 -(\frac{\eta^2}{4\epsilon}) 
\frac{\lambda}{4\pi N} \right ] + X^{\delta}(\eta,\mu)
+\frac{T^2}{4\pi^2} J^2_2(\eta/T,\mu/T)
\;. 
\label{AVdelta1} 
\end{eqnarray} 
The remaining divergent contributions come from purely vacuum (field
independent) graphs and can be absorbed by the zero point energy subtraction
counterterm

\begin{equation} 
X^{\delta}(\eta,\mu)= \frac{\eta^2}{2\pi \epsilon} -
\frac{\delta\:\eta^2}{\pi \epsilon} 
\left[1 -\frac{\lambda}{4\pi N}(\frac{1}{4\epsilon}) \right ] \,\,\,,
\end{equation} 
{}finally giving the finite effective potential

\begin{eqnarray} 
\frac {V_{{\rm eff},\delta^1}}{N} (\sigma_c, \eta, T, \mu) &=&
\delta \frac {\sigma_c^2}{2 \lambda} - 
 \frac{1}{2\pi} \left \{ \eta^2 \left [ \frac {1}{2} + \ln \left ( 
\frac {M}{\eta} \right ) \right ] + 2 T^2 I_1(\eta/T,\mu/T) \right \} 
\nonumber \\ 
&+& \delta \frac{\eta(\eta-\sigma_c)}{\pi} 
\left[\ln\left(\frac{M}{\eta}\right) - I_2 (\eta/T,\mu/T) \right] 
\nonumber \\ 
&+& \delta \frac {\lambda}{4\pi^2 N} \left\{ \eta^2  
  \left [ \ln \left ( \frac {M}{\eta}  \right ) - I_2(\eta/T,\mu/T) 
\right ]^2 +T^2 J^2_2(\eta/T,\mu/T) \right\}\;, 
\label{AVdelta1_2} 
\end{eqnarray} 
which is the result shown in Eq. (\ref{Vdelta1}). More details about
renormalization within the LDE can be found in the first work of 
Ref. \cite{prd1} (see also \cite{ldegn,jldamien} for the GN model).

\end{document}